\documentclass[fleqn,11pt,twoside]{article}

\usepackage{amsthm,amsthm,amssymb, color, xcolor,epsfig, graphics, subfigure}

\usepackage{amsmath, graphicx, latexsym, lscape}
\usepackage{cite}
\usepackage{caption}
\usepackage{xcolor}
\usepackage{dutchcal}
\usepackage{float}
\usepackage{hyperref}
\makeatletter

\def\bea{\begin{eqnarray}}
\def\eea{\end{eqnarray}}

\newtheorem{thm}{Theorem}[section]

\newtheorem{propn}[thm]{Proposition}

\newtheorem{rem}[thm]{Remark}

\newtheorem{cor}[thm]{Corollary}
\def\te{\tilde{\epsilon}}
\def\bea{\begin{eqnarray}}
\def\eea{\end{eqnarray}}

\newcommand{\sgn}{\text{sgn}}

\newcommand{\copyrightnote}[2]{{\renewcommand{\thefootnote}{}
 \footnotetext{\small\it
\begin{flushleft}
 \copyright \ #1   #2  
\end{flushleft}}}}

\newcommand{\Name}[1]{\begin{flushleft}
                       \LARGE \bf #1
                       \end{flushleft}\vspace{-3mm}}

\newcommand{\Author}[1]{\begin{flushleft}
                       \it #1 \end{flushleft}}

\newcommand{\Address}[1]{\begin{flushleft}
                       \it #1 \end{flushleft}}

\newcommand{\Date}[1]{\begin{flushleft}
                      \small  \it #1 \end{flushleft}}

\newcommand{\evenhead}{Author \ name}
\newcommand{\oddhead}{Article \ name}

\renewcommand{\@evenhead}{
\hspace*{-3pt}\raisebox{-15pt}[\headheight][0pt]{\vbox{\hbox to \textwidth
{\thepage \hfil \evenhead}\vskip4pt \hrule}}}
\renewcommand{\@oddhead}{
\hspace*{-3pt}\raisebox{-15pt}[\headheight][0pt]{\vbox{\hbox to \textwidth
{\oddhead \hfil \thepage}\vskip4pt\hrule}}}
\renewcommand{\@evenfoot}{}
\renewcommand{\@oddfoot}{}

\long\def\@makecaption#1#2{%
  \vskip\abovecaptionskip
  \sbox\@tempboxa{\small \textbf{#1.}\ \ #2}%
  \ifdim \wd\@tempboxa >\hsize
    {\small \textbf{#1.}\ \ #2}\par
  \else
    \global \@minipagefalse
    \hb@xt@\hsize{\hfil\box\@tempboxa\hfil}%
  \fi
  \vskip\belowcaptionskip}

\newcommand{\JNMPnumberwithin}[3][\arabic]{%
  \@ifundefined{c@#2}{\@nocounterr{#2}}{%
    \@ifundefined{c@#3}{\@nocnterr{#3}}{%
      \@addtoreset{#2}{#3}%
      \@xp\xdef\csname the#2\endcsname{%
        \@xp\@nx\csname the#3\endcsname .\@nx#1{#2}}}}%
}

\newcommand{\resetfootnoterule} {
  \renewcommand\footnoterule{%
  \kern-3\p@
  \hrule\@width.4\columnwidth
  \kern2.6\p@}
}

\renewcommand{\footnoterule}{}

\makeatother

\theoremstyle{definition}

\begin{document}

\renewcommand{\evenhead}{ {\LARGE\textcolor{blue!10!black!40!green}{{\sf \ \ \ ]ocnmp[}}}\strut\hfill 
O Ragnisco and F Zullo
}
\renewcommand{\oddhead}{ {\LARGE\textcolor{blue!10!black!40!green}{{\sf ]ocnmp[}}}\ \ \ \ \   
The integrable Volterra system and infinitely many species
}

\thispagestyle{empty}
\newcommand{\FistPageHead}[3]{
\begin{flushleft}
\raisebox{8mm}[0pt][0pt]
{\footnotesize \sf
\parbox{150mm}{{Open Communications in Nonlinear Mathematical Physics}\ \  \ {\LARGE\textcolor{blue!10!black!40!green}{]ocnmp[}}
\ \ Vol.6 (2026) pp
#2\hfill {\sc #3}}}\vspace{-13mm}
\end{flushleft}}

\FistPageHead{1}{\pageref{firstpage}--\pageref{lastpage}}{ \ \ Article}

\strut\hfill

\strut\hfill

\copyrightnote{The author(s). Distributed under a Creative Commons Attribution 4.0 International License}

\Name{The integrable Volterra system in the case of infinitely many species, either countable or uncountable.}

\Author{O. Ragnisco$^a$, F. Zullo$^b$}

\Address{$^a$Dipartimento di Matematica e Fisica, Universit\`a degli Studi ``Roma TRE" (retired)\\
Via della Vasca Navale 84,00146 Roma\\
e-mail: oragnisco@gmail.com\\ $^b$ DICATAM, Universit\`a degli Studi di Brescia
\\via Branze, 38 - 25123 Brescia, Italy \&\\
INFN, Milano Bicocca\\
Piazza della Scienza 3, Milano, 20126, Italy\\
e-mail: federico.zullo@unibs.it}

\Date{Received January 22, 2026; Accepted February 22, 2026}

\setcounter{equation}{0}

\smallskip

\noindent
{\bf Citation format for this Article:}\newline
 Orlando Ragnisco and Federico Zullo, 
 The integrable Volterra system in the case of infinitely many species, either countable or uncountable,
{\it Open Commun. Nonlinear Math. Phys.}, {\bf 6}, ocnmp:17363, \pageref{firstpage}--\pageref{lastpage}, 2026.

\strut\hfill

\noindent
{\bf The permanent Digital Object Identifier (DOI) for this Article:}\newline
{\it 10.46298/ocnmp.17363}

\strut\hfill

\begin{abstract}
\noindent 
In the present paper we derive a further extension of the results contained in two recent articles, both published in Open Communications in Nonlinear Mathematical Physics, where it was shown that the integrable version of the $N-$species Volterra model, introduced by V. Volterra in 1937, is in fact maximally superintegrable. Here we point out that the superintegrability property applies as well to the case of infinitely many competing species, either countable or uncountable. Analytical and numerical results are given.
\end{abstract}

\label{firstpage}


\section{Introduction}
The interest on dynamical systems of Lotka-Volterra type has been constantly growing all over the last century,  starting from the pioneering results contained in the original articles \cite{L1,L2, V1,Volterra1937}. Among  the extremely rich literature on the subject (see for instance \cite{Baigent9,CDE15,Bountis11,Hone17,fernandez6,Goodwin12,Griffin7,Hofba14,Peixe8,Leigh10,VanDerKamp13,Scalia16}), one paper deserves in our opinion a very special attention. It is the long essay published by Vito Volterra in 1937 \cite{Volterra1937},  where some results of capital importance have been achieved: first,  the original {\emph{predator-prey}} model  has been extended to an arbitrary number of species; second, the $N-$species model has been framed within   the context of Lagrangian and Hamiltonian mechanics; finally, a completely integrable subcase has been identified. Relying on  these outstanding results, the authors of the present paper performed some further hopefully non-trivial steps: namely, they discovered that the integrable version introduced in \cite{Volterra1937} was in fact maximally superintegrable \cite{RZ,SRTZ}, and as such reducible to a Hamiltonian system with only one degree of freedom. Here we make yet a further step, investigating  the extension to infinitely many competing species, either countable or uncountable,  always staying within the integrable Hamiltonian framework, and discover that the maximal superintegrability property is preserved by such generalization. Various different cases of the interaction operator are discussed, and a few periodic numerical examples are displayed. We may thus assert that {\emph{Infinity collapses to One}}. Let us anticipate that, with a suitable choice of the interaction parameters, the one dimensional system becomes the classical one body Hamiltonian $\mathcal{H}=\frac{P^2}{2}+V(Q)$.

To this aim, in Section 2  we briefly recall the results obtained in \cite{RZ}, \cite{SRTZ} for the integrable Volterra model in the case of a finite number of species. The new developments are contained in Sections 3 and 4. In particular, in Section 3 we consider the case of a denumerable collection of species, and in Section 4 we grapple with the problem of uncountable many species. Both Sections consist of different subsections, where we give first a general outlook and then face two different cases, concerning respectively even and odd  kernel of the interaction operators. In section 5 we first present and analyze some analytical examples  and then discuss
 numerical implementations. In Section 6 we recapitulate and critically analyze the results obtained and point out some affordable ways to improve the description of the system.

\noindent
\section{The case of a finite number N of species}

\noindent
The equations for the integrable $N$-species Volterra system have been introduced by V.Volterra in \cite{Volterra1937}, and thoroughly investigated in \cite{RZ,SRTZ}:
\begin{equation}
\frac{d N_r}{dt} = \epsilon_r N_r + \sum_{s\ne r=1}^N A_{rs}N_rN_s \quad\quad r=1,\ldots,N\label{model}.
\end{equation}
In the equation (\ref{model}), $N_r$ is the numerosity of the $r^{th}$ species, $\epsilon_r$ are the natural growth coefficients of each species and $A_{rs} $ are interaction coefficients between species $r$ and species $s$ that account for the effects of encountering between two individuals. In the integrable $N-$species case, the $N\times N$ interaction matrix $A$ has the form \cite{SRTZ}: 
\begin{equation}
A_{rs}=\epsilon_r\epsilon_s(B_r-B_s),\label{matrix}
\end{equation}
\noindent
where all the $ \epsilon_r$ and the $B_r$ are real non-zero numbers, and moreover $B_r\neq B_s$ if $r\neq s$. The matrix $A$ has always rank $2$, so  ${\mathcal I}$, the Image of $A$, is $2$-dimensional and  ${\mathcal K}$, its Kernel, is $(N-2)$-dimensional.
 Moreover ${\mathcal I}$ and ${\mathcal K}$ are two orthogonal subspaces with respect to the inner product $(u,v)\equiv \sum_j u_j v_j$ and the $N$-dimensional Euclidean space ${\mathbb E}_N$ can be written as the 
direct sum of them:
$${\mathbb E}_N = {\mathcal I} \oplus {\mathcal K}.$$
Also, it has been shown in \cite{SRTZ} that ${\mathcal I}$ is the linear span of two vectors, which are linearly independent because the coefficients $B_r$ are all distinct, namely $ \epsilon$ and $\eta$, of components $\epsilon_r$ and $\eta_r=B_r\epsilon_r$. 
The Kernel ${\mathcal K}$ is naturally defined as the $(N-2)$-dimensional subspace of 
$\mathbb{E}_N$ orthogonal to ${\mathcal I}$ and is such that its generic vector $k$ fulfils the linear  conditions 
\begin{equation}\label{kerA}
(k,\epsilon)=(k,\eta)=0.
\end{equation}
\noindent
On the other hand, the affine variety of equilibrium configurations  ${\mathcal E}$ is the  set  of elements  $z$ fulfilling
\begin{equation}\label{Z}
(z,\epsilon)=0,~ (z,\eta)=1.
\end{equation}

\noindent
It is convenient to rewrite the evolution equations (\ref{model}) In terms of the logarithmic variables defined by $y_j := \log N_j$, getting:
\begin{equation}
\dot y_j= \epsilon_j +\sum_{l=1}^N A_{jl}\exp(y_l) = \epsilon_j +( \eta_j\sum_{l=1}^N\epsilon_l-\epsilon_j\sum_{l=1}^N\eta_l)\exp(y_l)\label{llog}
\end{equation}

\noindent
It is possible to represent any vector $ \in {\mathbb E}_N$ as a linear combination of a vector $\in {\mathcal I}$ and a vector $\in {\mathcal K}$ as follows:
\begin{equation}
y_n=P\epsilon_n+Q\eta_n+ k_n \label{decompoy},
\end{equation}
\noindent
where in (\ref{decompoy})  $k_n$ is a generic element $k \in {\mathcal K}$.\footnote{To be precise, with the term $k_n$ we mean $\sum_{k=1}^{N-2}R_k\tau_m^{(k)}$, i.e. the decomposition of $k_n$ along a given orthonormal basis. We adopt here this shorter symbol for easiness of notation.}
\noindent
 The evolution equations (\ref{llog}) take the form:
\begin{equation}
\dot P \epsilon_n+\dot Q \eta_n + \dot{k}_n= \epsilon_n+\eta_n\sum_{k=1}^N \epsilon_k\exp(y_k) -\epsilon_n\sum_{=1}^N \eta_k\exp(y_k)
\end{equation}
\noindent
entailing:
\begin{eqnarray}
\dot P= 1-\sum_{n=1}^N\eta_n \exp(P\epsilon_n+Q\eta_n+ k_n),\label{eveq2a}\\
\dot Q = \sum_{n=1}^N\epsilon_n\exp(P\epsilon_n+Q\eta_n+ k_n), \label{eveq2b}\\
\dot{k}_n =0, \quad n=1...N-2\label{eveq2c}
\end{eqnarray}

\noindent
we can rewrite the Hamilton  equations in the canonical form:
\begin{equation}
\dot P= -\frac{\partial {\mathcal H}}{\partial Q}; ~~\dot Q = \frac{\partial {\mathcal H}}{\partial P}
\end{equation}
with the Hamiltonian explicitly given by:
\begin{equation}\label{h1}
{\mathcal H}=\sum_{n=1}^N \exp(P\epsilon_n+Q\eta_n+k_n) -Q.
\end{equation}
For other properties of this dynamical system, like its Poisson structure, the corresponding associated $N-2$ Casimirs, existence of equilibrium positions, nature of the orbits and some numerical examples we refer the reader to \cite{SRTZ}. In the next Section we are going to extend the above construction to the infinite chain, i.e. we let the numerosity of the species to be infinite.

\section{About the discrete infinite case}\label{sec3}
\subsection{General outlook}
In the discrete infinite case we  extend the involved summation to the whole discrete line, taking care of convergence properties. Here we have to point out that, on one hand, we can look at the operator $iA$ as a self-adjoint separable Hilbert-Schmidt operator  \cite{KF} mapping the Hilbert space $\ell^2$ of square summable sequences into itself, thanks to the  the inner product $(x,y) \doteq \sum_{n\in{\mathbb Z}}x_n y_n$. As such, its spectrum consists of two opposite real eigenvalues, say $\pm \omega$, and a null eigenvalue of infinite multiplicity, the Hilbert space $\ell^2$ being  spanned by the two orthonormal eigenvectors associated to $\pm \omega$ and by the orthogonal  subspace Ker$(A)$ of codimension 2. On the other hand,  $A$ acts nonlinearly on the space of sequences, according to the law $\dot y = A\exp (y)$, mapping $\ell^\infty$ into $\ell^2$. To be on the safe side, from now on we will assume that $\epsilon$ belongs to $\ell^2$, while $\eta$ and $\kappa$ belong to the dense open subset $S^\infty \subset \ell^2$ of rapidly decreasing sequences $s_n$ that vanish at $\infty$ faster than any power of $1/n$, i.e. such that $\lim_{n\to \infty}n^p s_n =0, \forall p$; this condition is quite natural for $\eta$, whose elements are defined by $\eta_n= B_n\epsilon_n$, and simply amounts to require that $B_n$ itself shares the same properties; its extension to $\kappa$ ensures that the whole quantity $Q\eta_n + \kappa_n$ enjoys the same regularity properties. Anyway, the possibility for $\eta$ and $\kappa$ to belong to other spaces will be also discussed in the next Sections.

The solution procedure is much the same as the one holding for finite $N$. We report the main formulas. The image of the summation operator
$$(Af)_n \doteq\sum_{m \in {\mathbb Z}}(\epsilon_m \eta_n-\epsilon_n \eta_m)f_m$$
\noindent
is the linear span of $\{\epsilon_n \},\{\eta_n \}$, to be considered as $\ell^2$ sequences;  as previously mentioned, $f_n=\exp(y_n)$ belong to $\ell^\infty$. In terms of the logarithmic variable $y_n$, the integrable Volterra system takes the following form:
\begin{equation}
\dot y_n = \epsilon_n + \sum_{m\in {\mathbb Z}}( \epsilon_m\eta_n -\eta_m \epsilon_n) \exp (y_m).\label{Voltdis}
\end{equation}
The sequence $y_n$ can be projected on the two complementary  subspaces Im$(A)$ and Ker$(A)$ in the form:
\begin{equation}
y_n= P\epsilon_n + Q\eta_n + \kappa_n\label{project}
\end{equation}
where $\kappa_n$  just depends on the initial conditions, belonging to the infinite dimensional subspace ${\mathcal K}$ of codimension $2$ defined by the linear equations:
\begin{equation}
\sum_{n\in {\mathbb Z}}k_n\epsilon_n=0, \quad \sum_{n\in {\mathbb  Z}}\kappa_n\eta_n=0.\label{ker}
\end{equation}
Since $\epsilon_n,\eta_n$ and $\kappa_n$ both converge to zero as $|n| \to \infty$, the numerosity $N_n \to 1$ as $|n|\to \infty$. The equations (\ref{Voltdis}) split in the following $3$ equations:
\begin{eqnarray}\label{eqm}
\dot P=1 - \sum_{m\in {\mathbb Z}}\eta_m\exp(P\epsilon_m+Q\eta_m+\kappa_m)\label{Pdiscr}\\
\dot Q=  \sum_{m\in {\mathbb Z}}\epsilon_m\exp(P\epsilon_m+Q\eta_m+\kappa_m)\label{Qdiscr}\\
\dot \kappa_n=0\label{kdiscr}
\end{eqnarray}
As stated before the sequences $\eta_n, \kappa_n$  are assumed to vanish at infinity faster than any power of $1/n$. As we shall see, in  view of some concrete examples that we will consider, one can also just assume that $\eta_n,\kappa_n$ vanish  at infinity exponentially or faster.

\noindent
The equilibrium position, if any, is a pair $P^0,Q^0$ such that:
\begin{eqnarray}
0=1-\sum_{m\in {\mathbb Z}}\eta_m\exp(P^0\epsilon_m+Q^0\eta_m+\kappa_m)\label{eqdiscrp}\\
0= \sum_{m\in {\mathbb Z}}\epsilon_m\exp(P^0\epsilon_m+Q^0\eta_m+\kappa_m)\label{eqdiscrq}
\end{eqnarray}
The sequence $z_n \doteq \exp(P^0\epsilon_n+Q^0\eta_n+k_n)\doteq z_n(P^0,Q^0)$ is the {\it equilibrium sequence}. Under the previous conditions imposed to $\epsilon_n,~ \eta_n,~ k_n$ the equilibrium sequence will belong to $\ell^\infty$.

Should we follow the alternative way of the spectral decomposition of the operator $A$ and apply the Hilbert-Schmidt machinery, as we did in \cite{RZ,SRTZ} for the case of finite species, we had  to assume that both the sequences  $\epsilon_n$ and $\eta_n$ belong to $\ell^2$.
 We  would  have to look at the roots of the secular equation:

\begin{equation}
\lambda_\pm v^\pm (n)= \alpha^\pm\eta_n-\beta^\pm\epsilon_n\label{discrteig}
\end{equation}

\noindent
As a matter of fact, the non-zero eigenvalues read:
\begin{equation}
\lambda_\pm \doteq \pm i\omega = \pm i{\sqrt {|\epsilon|^2 |\eta|^2-(\epsilon,\eta)^2}}\label{eigenv}
\end{equation}
\noindent
where we used the shorthand notations:
$$ |f|^2= \sum_{n\in {\mathbb Z}}f_n^2; \quad (f,g)= \sum_{n\in {\mathbb Z}} f_ng_n$$

\noindent
Notice that $\omega$, i.e. the expression under the square root, is strictly  positive because of Cauchy-Schwartz inequality. The equations for the eigenfunctions are given by:
\begin{eqnarray}
|\eta|^2\alpha-(\eta,\epsilon)\beta = \lambda \beta\label{deigeneq1}\\
(\epsilon,\eta)\alpha - |\epsilon|^2\beta=\lambda \alpha\label{deigeneq2}
\end{eqnarray}

 \noindent
By solving for $\alpha,\beta$ the system (\ref{deigeneq1}) and (\ref{deigeneq2})  in terms of variables $\epsilon,\eta$ we get for the eigenfunctions
$v_n^\pm \doteq u_n\pm i w_n$ the following expressions:

\begin{equation}\label{uv}
u_n=\frac{\epsilon_n}{|\epsilon |} , \quad  w_n=\frac{1}{|\epsilon|\omega}\left((\epsilon,\eta)\epsilon_n- |\epsilon|^2\eta_n\right)
\end{equation}

\noindent
The sequences $u_n,w_n$ are $\ell^2$ sequences with norm $1$ and are mutually orthogonal, namely $\sum_{n\in {\mathbb Z}} u_n w_n=0$. The interaction matrix $A_{n,m}$ now takes the form 
\begin{equation}
A_{n,m}=i\omega(w_mu_n-w_nu_m).
\end{equation}
The projection of $y_n$, equivalent to (\ref{project}), now reads
\begin{equation}
y_n= pu_n - qw_n + \kappa_n\label{project1},
\end{equation}
whereas the equations of motion (\ref{Voltdis}) reduce to
\begin{eqnarray}
\dot p= |\epsilon|+\omega\sum_{m \in\mathbb{Z}} w_m \exp(pu_m - qw_m + \kappa_m),\label{ev2a}\\
\dot q = -\omega\sum_{m \in\mathbb{Z}} u_m \exp(pu_m - qw_m + \kappa_m), \label{ev2b}\\
\dot k_m =0, 
\end{eqnarray}
The corresponding Hamiltonian is then given by
\begin{equation}
\mathcal{H}=\omega\left(-|\epsilon|q+\sum_{m \in\mathbb{Z}} \exp(pu_m - qw_m + \kappa_m) \right)
\end{equation}
In the next Section we will show how it is possible to give some very general expressions for the equations of motion in the infinite discrete case. This is somewhat surprising since in the case of finite species we have not been able to achieve such general results. In the infinite case we have been constrained to choose suitable parameters in order to achieve the convergence and this constraint led to the general expressions that we will give. In the Conclusions these aspects will be further commented.
\noindent
By looking at equations (\ref{ker}), in order to be operative it is natural to assume that the sequences $\epsilon_n$ and $\eta_n$ have an opposite parity with respect to the Kernel sequence $\kappa_n$. So, in the following two subsections, we will consider the case of a Kernel given by an even sequence (and hence $\eta_n$ and $\epsilon_n$ odd sequences), and the case of a Kernel given by an odd sequence (and hence $\eta_n$ and $\epsilon_n$ even sequences). From our point of view this is not restrictive: indeed we will show that it is possible to consider a mixed Hamiltonian flow that contains both the even functions and the odd ones. The conditions to get the convergence will be specified in the corresponding subsections.

\subsection{The case of an even Kernel}

We define the following sequences for $\epsilon_n$, $\eta_n$ and $\kappa_n$:
\begin{equation}\label{cho1}
\begin{split}
& \epsilon_n=\frac{\tilde{\epsilon}_n}{\sqrt{2N}}, \quad |\te_n|^2=1 \quad \te_{-n}=-\te_{n},\\
&\eta_n=-\eta_{-n}, \\
&\kappa_n=\kappa_{-n}
\end{split}
\end{equation}
In the previous it is assumed that $\te_n$ has compact support in $[-N,N]$, so it is an odd function  of $n$ assuming the values $\pm 1$ on the interval $[-N, N]$. The sequence $\eta_n$ is also odd, whereas $\kappa_n$ is even since it must be orthogonal to the sequences $\epsilon_n$ and $\eta_n$. Notice that $\epsilon_n$ is normalized in  $\ell^2$ to 1, and the same normalization can be chosen for $\eta_n$. As previously mentioned, both $\eta_n$ and $\kappa_n$ are assumed to belong to the dense open subset $S^\infty$ of $\ell^2$ consisting in rapidly decreasing sequences. 
We start with the equation for $\dot{Q}$:
\begin{equation}\label{Qeq} 
\dot{Q}=\sum_{n\in {\mathbb Z}}\epsilon_n\exp(P\epsilon_n+Q\eta_n+\kappa_n).
\end{equation}

In order to deal with equation (\ref{Qeq}), we make use of the following
\begin{propn}\label{prop1}
Let  $s_n$ be a sequence in  $S^\infty$ and consider the following expression:
\begin{equation}\label{SP}
S(P,x,N)=\sum_{n=-N}^N \epsilon_n\exp(P\epsilon_n)f(xs_n),
\end{equation}
where $f(x)$ is a continuous function of $x$ and $\epsilon_n$ is given in (\ref{cho1}). Then, for any finite values of $x$ one has
\begin{equation}
\lim_{N\to\infty} S(P,x,N)=Pf(0).
\end{equation}
\end{propn}
\textbf{Proof}. By taking the Taylor expansion of the exponential we get
\begin{equation}
S(P,x,N)=\sum_{n=-N}^{N}\epsilon_n(1+P\epsilon_n+\frac{P^2}{2}\epsilon_n^2+\ldots)f(x s_n).
\end{equation} 
Let us analyze the term proportional to $\epsilon_n^2$. It is given by
\begin{equation}
P\sum_{n=-N}^{N}\epsilon_n^2f(x s_n)=P\frac{1}{2N}\sum_{n=-N}^Nf(x s_n).
\end{equation}
The previous is the arithmetic mean of $f(x s_n)$, so we can use the Cauchy's limit theorem on the arithmetic mean of sequences: since $s_n$ converges to $0$ as $n \to \infty$ and $f(x)$ is continuous we get in the limit $N \to \infty$
\begin{equation}
\lim_{N\to\infty} P\sum_{n=-N}^{N}\epsilon_n^2f(x s_n)=\lim_{N\to\infty}P\frac{1}{2N}\sum_{n=-N}^Nf(x s_n)=Pf(0).
\end{equation}
It follows that all the powers of $\epsilon$ greater than $2$ do not give a contribution in the Taylor series expansion of $\exp(P\epsilon_n)$ in (\ref{SP}). 

  Let us consider the contribution from the term linear in $\epsilon_n$. 
\begin{equation}
\sum_{n=-N}^{N}\epsilon_n f(xs_n)=\frac{1}{\sqrt{2N}}\sum_{n=1}^N \te_n\left(f(x s_n)-f(x s_{-n})\right)
\end{equation}
Notice that, since $s_n$ is  a sequence in  $S^\infty$,   $f(xs_n)$ is a bounded sequence and in the limit $n \to \infty$ (and hence $N\to \infty$) it converges to $f(0)$ (the sequence $f(xs_n)-f(0)$ is bounded and approaches zero). The above sum, by adding and subtracting $f(0)$, can be rewritten as
\begin{equation}
\sum_{n=-N}^{N}\epsilon_n f(xs_n)=\sum_{n=1}^N \epsilon_n\left(f(x s_n)-f(0)\right)-\sum_{n=1}^N \epsilon_n\left(f(x s_{-n})-f(0)\right)
\end{equation}
and both the addenda converge to zero in the limit $N \to \infty$. For example for the first addendum  on the right hand side, if $f(x)$ is Lipschitz at $x=0$, i.e. if $|f(xs_n)-f(0)|\leq C|xs_n|$, one has 
\begin{equation}
|\frac{1}{\sqrt{2N}}\sum_{n=1}^N \te_n\left(f(x s_n)-f(0)\right)|\leq \frac{C|x|}{\sqrt{2N}}\sum_{n=1}^{N} |s_n|
\end{equation}
and, since $s_n \in S^\infty$ the sum is finite and in the limit $N\to \infty$ the linear term in $\epsilon_n$ in the Taylor series expansion of $\exp(P\epsilon_n)$ in (\ref{SP}) vanishes. $\square$

The fact that $\epsilon_n$ is an odd sequence is crucial in the vanishing of the linear term in $\epsilon_n$. As a result of Proposition (\ref{prop1}) we immediately get the following
\begin{cor}
The equation of motion (\ref{Qeq}) reduces, in the limit $N\to\infty$ to
\begin{equation}\label{dotqodd}
\dot{Q}=P.
\end{equation}
\end{cor}
Indeed, it is sufficient to take $f(xs_n)=\exp(Q\eta_n+\kappa_n)$ since both $\eta_n$ and $\kappa_n$ belong to $S^\infty$.

Now we look at the equation for $\dot{P}$.  One has
\begin{equation}\label{dpeq}
\dot{P}=1-\sum_{n\in {\mathbb Z}}\eta_n\exp(P\epsilon_n+Q\eta_n+\kappa_n).
\end{equation}
In order to deal with the sum, we introduce the following
\begin{propn}\label{prop2}
Let $\eta_n$ and  $s_n$ be two sequences in $S^\infty$ and consider the following expression:
\begin{equation}\label{SP2}
T(P,x,N)=\sum_{n=-N}^N \eta_n\exp(P\epsilon_n)f(xs_n),
\end{equation}
where $f(x)$ is a continuous function of $x$ and $\epsilon_n$ is given in (\ref{cho1}). Then, for any finite values of $x$ one has
\begin{equation}\label{lim1}
\lim_{N\to\infty} T(P,x,N)=\sum_{n\in \mathbb{Z}} \eta_n f(xs_n)
\end{equation}
\end{propn}
\textbf{Proof}. Indeed, if one consider the sum $\sum_{n=-N}^N \eta_n f(xs_n)$, it is convergent in the limit $N\to \infty$. For large $n$ the terms $s_n \to 0$ so $f(xs_n)\to f(0)$. Being continuous, $f(x)$ is locally bounded around $x=0$ so we can find an $M$ such that $|f(x s_n)|<M$ for all values of $n$. It follows the convergence of the sum $\sum_{n=-N}^N \eta_n f(xs_n)$ from the $S^\infty$ convergence of $\eta_n$. If one considers the Taylor expansion of $\exp(P\epsilon_n)$ in the sum (\ref{SP2}), only the constant term survives in the limit $N\to \infty$ and we get (\ref{lim1}). $\square$

By using Proposition (\ref{prop2}) we get for the expression (\ref{dpeq})
\begin{equation}\label{dotpodd}
\dot{P}=1-\sum_{n\in {\mathbb Z}}\eta_n\exp(Q\eta_n+\kappa_n).
\end{equation}
The previous expression can be written in a different form by using the fact that $\eta_n$ is an odd function and $\kappa_n$ even. We get
\begin{equation}\begin{split}
&\sum_{n\in {\mathbb Z}}\eta_n\exp(Q\eta_n+\kappa_n)=\sum_{n=1}\eta_n\exp(Q\eta_n+\kappa_n)-\sum_{n=1}\eta_n\exp(-Q\eta_n+\kappa_n)=\\
&=2\sum_{n=1}\eta_n\exp(\kappa_n)\sinh(Q\eta_n)=\sum_{n\in\mathbb{Z}}\eta_n\exp(\kappa_n)\sinh(Q\eta_n)
\end{split}
\end{equation}
giving
\begin{equation}\label{dotpodd1}
\dot{P}=1-\sum_{n\in\mathbb{Z}}\eta_n\exp(\kappa_n)\sinh(Q\eta_n).
\end{equation}
The corresponding Hamiltonian reads
\begin{equation}\label{evenH}
\mathcal{H}=\frac{P^2}{2}-Q+\sum_{n\in\mathbb{Z}}\exp(\kappa_n)\left(\cosh(Q\eta_n)-1\right).
\end{equation}
Notice that we subtracted the constant term $-1$ to each term of the sequence in the Hamiltonian in order to ensure the convergence. The same situation occurs in the infinite Toda lattice \cite{Toda,Teschl} and in the integrable infinite Volterra lattice obtained from the
infinite Toda by the reduction $B_n=0$ \cite{KVM,Toda}. This is the analogous to the renormalization procedure in field theory when the vacuum state energy is subtracted to the Hamiltonian.

\subsection{The case of an odd Kernel}
Now we specify the sequences $\epsilon_n$, $\eta_n$ and $\kappa_n$ as:
\begin{equation}\label{cho2}
\begin{split}
& \epsilon_n=\frac{\tilde{\epsilon}_n}{\sqrt{2N}}, \quad |\te_n|^2=1 \quad \te_{-n}=\te_{n},\\
&\eta_n=\eta_{-n},\\
&\kappa_n=-\kappa_{-n}
\end{split}
\end{equation}
In the previous again it is assumed that $\te_n$ has compact support in $[-N,N]$. We further assume that\footnote{It is known \cite{AS} that if the values of $\tilde{\epsilon}_n$ are chosen randomly, then $\sup_{\theta\in\mathbb{R}}|\sum_{n=0}^N\te_n e^{\textrm{i}n\theta}| =O(\sqrt{N\log(N)})$ with probability 1. Other interesting sequences that can be considered are the automatic sequences, like the Rudin-Shapiro sequence, for which one has $\sum_{n=0}^N\te_n =O(\sqrt{N})$ (see also \cite{AS}).}
\begin{equation}\label{con2}
\sum_{n=-N}^N \te_n= o(\sqrt{N}), \quad \textrm{i.e.} \quad \lim_{N\to\infty}\sum_{n=-N}^N\epsilon_n=0.
\end{equation}
This condition will ensure the convergence of the formulae for the equations of motion as we will see. 

Again, we will give two Propositions in order to obtain the equations of motion.
\begin{propn}\label{prop3}
Let  $s_n$ be a sequence in $S^\infty$ and consider the following expression:
\begin{equation}\label{MP}
M(P,x,N)=\sum_{n=-N}^N \epsilon_n\exp(P\epsilon_n)f(xs_n),
\end{equation}
where $f(x)$ is a continuous function of $x$ and $\epsilon_n$ is given in (\ref{cho2}) and satisfies (\ref{con2}). Then, for any finite values of $x$ one has
\begin{equation}
\lim_{N\to\infty} M(P,x,N)=Pf(0).
\end{equation}
\end{propn}
\textbf{Proof}. Let us take the Taylor series of the exponential in (\ref{MP}):
\begin{equation}
M(P,x,N)=\sum_{n=-N}^N \epsilon_n\left(1+P\epsilon_n+\frac{P^2}{2}\epsilon_n^2+\ldots\right)f(xs_n),
\end{equation}
and analyze firstly the term proportional to $\epsilon_n^2$:
\begin{equation}
P\sum_{n=-N}^{N}\epsilon_n^2f(x s_n)=\frac{1}{2N}\sum_{n=-N}^N f(x s_n).
\end{equation}
From the Cauchy's limit theorem on the arithmetic mean of sequences we see that this term converges to $Pf(0)$. It follows that all the powers of $\epsilon$ greater than $2$ do not give a contribution in the Taylor series expansion of $\exp(P\epsilon_n)$ in (\ref{MP}). It remains only the contribution from the term linear in $\epsilon_n$. One has
\begin{equation}\label{eqpu}
\sum_{n=-N}^{N}\epsilon_n f(xs_n)=\frac{1}{\sqrt{2N}}\sum_{n=-N}^N \te_n f(x s_n).
\end{equation}
Due to the condition (\ref{con2}), since $f(x s_n)$ is bounded, the sum on the right hand side is $o(\frac{1}{\sqrt{N}})$ and it follows that the contribution from the term linear in $\epsilon$ in (\ref{MP}) vanishes $\square$.

From the Proposition (\ref{prop3}) one has the following
\begin{cor}
The equation of motion for $Q$, i.e.
\begin{equation}
\dot{Q}=\sum_{n=-N}^{N}\epsilon_n \exp(P\epsilon_n+Q\eta_n+\kappa_n)
\end{equation}
reduces, in the limit $N\to \infty$, to
\begin{equation}
\dot{Q}=P.
\end{equation}
\end{cor}

\begin{figure}
\centering
\includegraphics[scale=0.5]{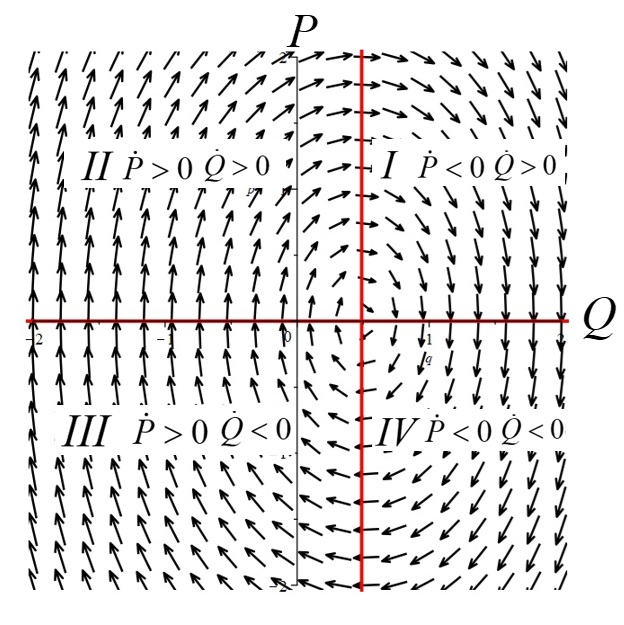}
\caption{Plot of the curves defined by $\dot{Q}=0$ and $\dot{P}=0$ (red) dividing the plane $(Q,P)$ in four regions, each having a definite sign for $\dot{Q}$ and $\dot{P}$.}
\label{fig1}
\end{figure}

Let us now look at the equation of motion for $P$. We notice that Proposition (\ref{prop2}) is still valid if $\epsilon_n$ is as given in (\ref{cho1}), the proof being the same. So we can directly write
\begin{equation}
\dot{P}=1-\sum_{n\in \mathbb{Z}}\eta_{n}\exp(Q\eta_n+\kappa_n).
\end{equation} 
Now $\eta_n$ is even and $\kappa_n$ is odd and since the interval of summation is symmetric only the even portion of $\exp(\kappa_n)$ survives, giving
\begin{equation}
\dot{P}=1-\sum_{n\in \mathbb{Z}}\eta_{n}\exp(Q\eta_n)\cosh(\kappa_n).
\end{equation} 
The corresponding Hamiltonian is given by
\begin{equation}\label{oddH}
\mathcal{H}=\frac{P^2}{2}-Q+\sum_{n\in \mathbb{Z}}\exp(Q\eta_n)\left(\cosh(\kappa_n)-1\right).
\end{equation}
Let us now give a Remark about the equilibrium positions.
\begin{rem}\label{rem1}
The equations of motion, both for an odd Kernel and for an even one, are given by
\begin{equation}\label{eqmotrem}
\dot{Q}=P, \quad \dot{P}=1-\sum_{n\in \mathbb{Z}}\eta_n\exp(Q\eta_n+\kappa_n),
\end{equation}
where the sequences $\epsilon_n$ and $\eta_n$ have opposite parity with respect to $\kappa_n$. The above system has always just one equilibrium position, say $(Q,P)=(Q_0,0)$ if the sequence $\eta_n$ has not a definite sign. In this case the motion is bounded and the equilibrium position is a center.
\end{rem}
Indeed the function defining $\dot{P}$ in (\ref{eqmotrem}) is a decreasing function of $Q$ as it follows immediately by taking its derivative with respect to $Q$. Further, for $Q\to -\infty$, $\dot{P} \to +\infty$ (we assume that there will be always at least one $\eta_n$ negative), whereas for $Q \to +\infty$, $\dot{P} \to -\infty$ (we assume that there will be always at least one $\eta_n$ positive). It follows that the equation $\dot{P}=0$ has just one zero. Further, the motion must be bounded and periodic. Notice that the equilibrium point is stable but not asymptotically stable: this easily follows from the linearization of the system (\ref{eqmotrem}) around the equilibrium point, that is a center. The fact that the motion is bounded and periodic follows from a global analysis of the orbits in the plane $(Q,P)$, or, more easily, by looking at the potential in the Hamiltonian defining the Hamiltonian system (\ref{eqmotrem}), given by 
\begin{equation}
\mathcal{H}=T+V=\frac{P^2}{2}-Q+\sum_{n\in\mathbb{Z}}\left(\exp(Q\eta_n+\kappa_n)-1\right).
\end{equation} 
The potential 
\begin{equation}
V=-Q+\sum_{n\in\mathbb{Z}}\left(\exp(Q\eta_n+\kappa_n)-1\right)
\end{equation} has one global minimum $Q=Q_0$.  Indeed, by assuming that the sequence $\eta_n$ has not a definite sign, $V$ approaches $+\infty$ for $Q\to \pm \infty$. Also, its derivative has a unique zero, as shown by Remark (\ref{rem1}). This zero is a minimum for the potential $V$, as can be easily seen by taking the second derivative of $V$. So $V(Q)$ is an infinite potential well and the motion is bounded and periodic. In Figure (\ref{fig1}) we report the nullclines corresponding to $\dot{Q}=0$ and $\dot{P}=0$ in red and the arrows displaying the direction of the flow in the four region of the plane $(Q,P)$ defined by the nullclines. 

\begin{rem}\label{rem1.1}
If the sequence $\eta_n$ has just one sign, then the system (\ref{eqmotrem}) has one equilibrium position if $\eta_n\geq 0$ and has no equilibrium position (open orbits) if $\eta_n \leq 0$.
\end{rem}
The previous Remark follows by looking at the properties of the system (\ref{eqmotrem}) when $\eta_n\geq 0$ or $\eta_n\leq 0$, as did for the Remark (\ref{rem1}). Indeed if $\eta_n\leq 0$ for all $n$ the equation defining $\dot{P}$, i.e.
\begin{equation}
\dot{P}=1-\sum_{n\in \mathbb{Z}}\eta_n\exp(Q\eta_n+\kappa_n),
\end{equation}
is always positive and $P$ increases monotonically in time, resulting in a unbounded motion. Vice versa, if $\eta_n\geq 0$ then the equation $\dot{P}=0$ has just one solution since $1-\sum_{n\in \mathbb{Z}}\eta_n\exp(Q\eta_n+\kappa_n)$ is a decreasing is a decreasing function of $Q$ and goes from $1$ for $Q\to -\infty$ to $-\infty$ for $Q \to +\infty$. The equilibrium position is again a center in this case and the orbits are bounded and periodic.

\begin{rem}\label{rem2}
The equations of motion (\ref{eqmotrem}) holds true also if the sequences $\eta_n$ and $\kappa_n$ belong to $\ell^1$. The Propositions (\ref{prop1}), (\ref{prop2}) and (\ref{prop3}) indeed hold also for $\eta_n$ and $\kappa_n$ belonging to $\ell^1$.
\end{rem}
An example of sequences in $\ell^1$ will be given in the subsection (\ref{subexd}). To end this Section, we give a Remark about the possibility to get a mixed odd/even Hamiltonian.
\begin{rem} It is possible to take a flow that is a combination of the Hamiltonian containing the even Kernel (\ref{evenH}), say $\mathcal{H}_e$, and the Hamiltonian containing the odd Kernel, say $\mathcal{H}_o$:
\begin{equation}
\alpha\mathcal{H}_e+(1-\alpha)\mathcal{H}_o=\frac{P^2}{2}-Q+\sum_{n\in\mathbb{Z}}\left(\alpha\exp(Q\eta_n^o+\kappa^e_n)+(1-\alpha)\exp(Q\eta_n^e+\kappa^o_n)-1\right).
\end{equation}
where the superscripts on the sequences remind if they are odd or even. If the sequences $\eta_n^o$ and $\eta_n^e$ have not a definite sign, the potential
\begin{equation}
V(Q)=-Q+\sum_{n\in\mathbb{Z}}\left(\alpha\exp(Q\eta_n^o+\kappa^e_n)+(1-\alpha)\exp(Q\eta_n^e+\kappa^o_n)-1\right)
\end{equation}
has just one minima for $\alpha \in [0,1]$ and is an infinite potential well, so that the corresponding motion is bounded and periodic. The proof is the same as after Remark (\ref{rem1}) and is omitted.
\end{rem}

\section{About the continuous case}
\subsection{General outlook}
In the continuum limit, the dynamical system

\begin{equation}
\dot y_j = \epsilon_j+ \sum_{jk}A_{jk}\exp(y_k)\label{log}
\end{equation}

\noindent
is replaced by the integro-differential equation for the variable $y(x,t)$:

\begin{equation}
\frac{\partial y}{\partial t}=\epsilon(x)- \int_{-\infty}^\infty dx^\prime(\epsilon(x)\eta(x^\prime)-\eta(x)\epsilon (x^\prime))\exp [y(x^\prime)]\label{cneq}
\end{equation}

\noindent
Clearly, the right hand side is a linear combination of two functions, $\epsilon(x)$ and $\eta(x)$. Since $\eta(x)=B(x)\epsilon(x)$, the two functions will  be linearly independent if $B^\prime (x) \ne 0$ (almost everywhere). 
Further conditions that will come out in the following are: 

1)There must be a finite or countably infinite set of points where $\epsilon(x)$  changes its sign; 

2) both $\epsilon(x)$ and $\eta(x)$ must be square-summable on the whole real line. 

We recall the inclusion relations for the spaces $L^1, L^2, L^\infty$. If the functions are defined on a set of finite Lebesgue measure we have $L^1\subset L^2 \subset L^\infty$. In the case of a support of infinite measure. like for instance a semi-line or the whole real line, the inclusion relations are reversed: $L^\infty \subset L^2 \subset L^1$.

Even in this case, for the $existence$ of the equations of motion one needs that the argument of the exponential be uniformly bounded on the whole real line, while $\epsilon (x), \eta(x)$ must be  absolutely integrable, and then square integrable, on the whole real line. Of course these requirements are "a fortiori" satisfied by $S^\infty$ functions.

\noindent
It is then natural to look at  $y(x,t)$ as a linear combination of $\epsilon(x),\eta(x)$ with time-dependent coefficients, that we will call again  $P$ and $Q$, plus a term  $\kappa(x)$ orthogonal  to $\epsilon (x)$ and $\eta(x)$:

\begin{equation}
y(x,t) = P(t)\epsilon(x) + Q(t)\eta(x) +\kappa(x)\label{decomposition}
\end{equation}

\noindent
Accordingly, we will have:

\begin{eqnarray}
\dot P= 1-\int_{-\infty}^\infty dx  \eta(x)\exp[P\epsilon(x)+Q\eta(x)+\kappa(x)]\label{dotp}\\
\dot Q= \int_{-\infty}^\infty dx \epsilon(x) \exp[P\epsilon(x)+Q\eta(x)+\kappa(x)]\label{dotq}
\end{eqnarray}
\noindent
 Equations (\ref{dotp},\ref{dotq}) can be written in canonical form, namely

\begin{eqnarray}
\dot P=-\frac {\partial H}{\partial Q}\label{hameq1}\\
\dot Q =\frac{\partial H}{\partial P}\label{hameq2}
\end{eqnarray}

\noindent
$H$ being the Hamiltonian:

\begin{equation}
H= -Q +\int_{-\infty}^\infty dx \exp(P\epsilon(x)+Q\eta(x)+\kappa(x))\label{ham}
\end{equation}

\noindent
As in the infinite discrete case, the Hamiltonian  (\ref{ham})    does not seem to exist, even if  $\kappa(x),~ \epsilon(x), ~\eta(x)$  are uniformly bounded and rapidly decreasing functions. However, we remark that, analogously to the infinite discrete case,  if the exponential is uniformly bounded  on the whole real line, then  the improper integral can be made convergent by just subtracting to the integrand its asymptotic limit, i.e. $1$. 
\noindent

\noindent
We look now  at the spectral decomposition of the integral operator $A$, whose action on a given function $f(x)$ is given by:
$$(Af)(x):=\int_{-\infty}^\infty dx^\prime [\epsilon(x)\eta(x^\prime)-\eta(x)\epsilon(x^\prime)]f(x^\prime)$$

\noindent
$A$ is  a skew-symmetric Hilbert-Schmidt operator since its kernel is square summable if so are the functions $\epsilon(x)$ and $\eta(x)$;  its kernel is in fact the kernel of the Poisson Bracket between two functionals $F[y],G[y]$:

$$\{F,G\}:=\int_{-\infty}^\infty dx \int_{-\infty}^\infty dx^\prime \frac{\delta F}{\delta y(x)}A(x.x^\prime)\frac{\delta G}{\delta y(x^\prime)}$$

\noindent
We can apply to $iA$ all the theorems holding for compact self-adjoint operators \cite{KF} on a Hilbert space. In particular,  it has no residual spectrum, its eigenvectors, including those associated with its kernel, form a complete (orthonormal) set in $L^2({\mathbb R})$, its discrete spectrum is either finite or given by a decreasing sequence accumulating to $0$, and its norm is a point of the spectrum. Indeed, we have more, because the kernel $A(x,x^\prime)$ is a {\it separable kernel of rank 2}, and consequently the operator $A$ has only two non-zero complex conjugated eigenvalues.

\noindent
It is remarkable that the  results for eigenvalues and eigenfunctions holding in the discrete case migrate with no essential modifications to the continuum case.

\noindent
Namely, the eigenvalue equation reads:
\begin{equation}
\psi^\pm(x,\omega)=(1/\lambda_\pm)(\eta(x)\alpha_\pm-\epsilon(x)\beta_\pm))\label{eigenve}
\end{equation}

\noindent
The non-zero eigenvalues are given by:
\begin{equation}
\lambda_\pm \doteq \pm i\omega = \pm i{\sqrt {|\epsilon|^2 |\eta|^2-(\epsilon,\eta)^2}}\label{eigenv}
\end{equation}
\noindent
where we used the shorthand notations:
$$|f|^2\doteq\int_{-\infty}^\infty dx |f(x)|^2,\quad (f,g)\doteq\int_{-\infty}^\infty dx f(x)g(x).$$
\noindent
Notice that $\omega$, i.e. the expression under the square root, is strictly  positive because of Cauchy-Schwartz inequality. 

\noindent
The coefficients defining the eigenfunctions have to satisfy the homogeneous system:
\begin{eqnarray}
|\eta|^2\alpha-(\eta,\epsilon)\beta = \lambda \beta\label{eigeneq1}\\
(\epsilon,\eta)\alpha - |\epsilon|^2\beta=\lambda \alpha\label{eigeneq2}
\end{eqnarray}

 \noindent
Once solved for $\alpha,\beta$ the system (\ref{eigeneq1}), (\ref{eigeneq2}),  in terms of variables $\epsilon,\eta$ we get for the eigenfunctions
$\psi_\pm (x.\omega) \doteq u(x,\omega)\pm i v(x,\omega)$ the following expressions:

\begin{eqnarray}
u (x,\omega)=\epsilon(x)/|\epsilon |\label{u}\\
w(x,\omega) =(\epsilon (x)(\epsilon,\eta)-\eta(x)|\epsilon |^2)/(\omega |\epsilon |)\label{v}
\end{eqnarray}

\noindent
As in the finite $N$ case, $u(x,\omega)$ and $w(x,\omega)$ have norm one and are mutually orthogonal. Accordingly, the kernel $A(x,x^\prime)$ of the integral operator $A$ can be written in the form:

\begin{equation}
A(x,x^\prime) = i\omega (u (x,\omega)w(x^\prime,\omega)-u (x^\prime,\omega)w(x,\omega))\label{kereigenf}
\end{equation}

\subsubsection{Linearization around the eventual equilibrium configuration.}
Now, let us assume  that the system enjoys an equilibrium configuration, namely that there exist at least one point $P^0,Q^0$ in the $P,Q$-plane such that
the function $z(x) \doteq \exp[P^0 \epsilon(x)+Q^0 \eta(x)+\kappa(x)]$ fulfils  the properties  $(z,\epsilon)=0$, $(z,\eta)=1$, and ask whether there exist small oscillations around that equilibrium configuration. 
To this aim, we set $P=P^0+ \delta P,\quad Q=Q^0+ \delta Q$, and consider a first order expansion of the exponential, setting  

$$\exp[P\epsilon(x)+Q\eta(x)+\kappa(x)]\simeq z(x)(1+\delta P \epsilon(x)+\delta Q\eta(x))$$

\noindent
Accordingly, we get the following linear system of differential equations for  the unknowns $\delta P, \delta Q$:
\begin{eqnarray}
\dot{\delta P}= -{\cal A}\delta P -{\cal B}\delta Q\label{so1},\\
\dot{\delta Q}= {\cal C}\delta P+{\cal A}\delta Q \label{so2}.
\end{eqnarray}
In the previous we have denoted:
\begin{eqnarray}
{\cal A}=\int_{-\infty}^\infty dx z(x)\epsilon(x)\eta(x)\label{ABC1}\\
{\cal B}= \int_{-\infty}^\infty dx z(x)\eta(x)^2\label{ABC2}\\
{\cal C}=\int_{-\infty}^\infty dx z(x)\epsilon(x)^2\label{ABC3}
\end{eqnarray}
\noindent
We notice that $z(x)$ can be interpreted as a weight function in the previous integrals.

The nature of the system (\ref{so1}-\ref{so2}) is fixed by the solution of the secular equation:

\begin{equation}
\lambda^2-{\cal A}^2+{\cal B}{\cal C}=0\label{sec}
\end{equation}

It easy to see that 
\begin{equation}
-{\cal A}^2+{\cal B}{\cal C}= (1/2)\int_{-\infty}^\infty dx\int_{-\infty}^\infty dx^\prime z(x)z(x^\prime)[\epsilon(x)\eta(x^\prime)-\eta(x)\epsilon(x^\prime)]^2\label{det}
\end{equation}
\noindent
where (\ref{det})  is the trace of the square of the matrix $M$
\begin{equation}
M=\begin{pmatrix}
-{\cal A}& -{\cal }B \\
{\cal C}  & {\cal A} \label{M}
\end{pmatrix}
\end{equation}

\noindent
hence, the linearized motion will be periodic around the centrum $(P^0,Q^0)$ with period 
$$\omega={\sqrt{-{\cal A}^2+{\cal B}{\cal C}}}.$$
\noindent
provided $z(x)$ exists as a positive real function a.e. on ${\mathbb R}$.
\noindent

In the next Section, in parallel with the infinite discrete case, we will give some general expressions for the equations of motion. Again, in order to be operative, we assume that the functions $\epsilon(x)$ and $\eta(x)$ have an opposite parity with respect to the function $\kappa(x)$. In the following subsections we will consider the case of a Kernel given by an even function, and hence $\epsilon(x)$ and $\eta(x)$ odd, and the case of a Kernel given by an odd function, and hence $\epsilon(x)$ and $\eta(x)$ even.

\subsection{The case of an even Kernel}
Let us specify the functions $\epsilon$ and $\eta$ as:
\begin{equation}\label{etaeps}
\epsilon(x)=\frac{1}{\sqrt{2L}}\tilde{\epsilon}(x), \quad \eta(x)=-\eta(-x)
\end{equation} 

\noindent
where $\tilde{\epsilon}(x)^2=1$ for $x\in[-L,L]$, $\epsilon(x)=0$ for $x\notin [-L,L]$, $\epsilon(x)$ odd and $\eta(x)$ odd. We assume that both $\epsilon$ and $\eta$ are square integrable and normalized to 1. 
As for the Kernel $\kappa(x)$, it is represented by square integrable even functions, compactly defined in an interval or defined on the whole real line. Let us start from the equation for  $\dot{Q}$. We have
\begin{equation}\label{qpp}
\dot Q= \int_{-\infty}^\infty dx \epsilon(x) \exp[P\epsilon(x)+Q\eta(x)+\kappa(x)]
\end{equation}
We give the following
\begin{propn}\label{procon}
Let  $s(\xi)$ be a bounded  rapidly decreasing function and consider the following expression:
\begin{equation}\label{SPc}
S(P,x,L)=\int_{-L}^L \epsilon(\xi)\exp(P\epsilon(\xi))f(xs(\xi))d\xi,
\end{equation}
where $f(x)$ is a continuous function of $x$ and $\epsilon(\xi)$ is given in (\ref{etaeps}). Then, for any finite values of $x$ one has
\begin{equation}
\lim_{L\to\infty} S(P,x,L)=Pf(0).
\end{equation}
\end{propn}
\textbf{Proof}. By taking the Taylor expansion of the exponential we get
\begin{equation}
S(P,x,L)=\int_{-L}^{L}\epsilon(\xi)(1+P\epsilon(\xi)+\frac{P^2}{2}\epsilon(\xi)+\ldots)f(x s(\xi))d\xi.
\end{equation} 
The term proportional to $\epsilon^2$ is given by
\begin{equation}\label{intf}
P\int_{-L}^L\epsilon(\xi)^2f(xs(\xi))d\xi=\frac{P}{2L}\int_{-L}^Lf(xs(\xi))d\xi
\end{equation}
Consider the following expression
\begin{equation}
F(x,L)=\frac{1}{2L}\int_{-L}^L\left(f(xs(\xi))-f(0)\right)d\xi
\end{equation}
If the integral is finite for $L\to \infty$, then in the limit $L\to\infty$ one has $\lim_{L\to\infty}F(x,L)=0$, hence
\begin{equation}
\lim_{L\to\infty}\frac{1}{2L}\int_{-L}^L f(xs(\xi))d\xi=f(0).
\end{equation}
If the integral diverges for $L\to\infty$, by the L'H\^{o}spital rule one has
\begin{equation}
\lim_{L\to\infty}\frac{1}{2L}\int_{-L}^L\left(f(xs(\xi))-f(0)\right)d\xi=\lim_{L\to\infty}\frac{f(xs(L))+f(xs(-L))-2f(0)}{2}=0,
\end{equation}
where in the last equation we use the fact that $s(\xi)\in S^\infty$. Put in another way, in the Taylor series of the integral of $f(xs(\xi))$ (\ref{intf}) around $x=0$, only the first term $f(0)$ gives a finite contribution, all the others approach zero in the limit $L\to \infty$. 

From the previous result it follows that all the powers of $\epsilon$ greater than two in the expansion of $\epsilon\exp(P\epsilon)$ in (\ref{SPc}) don't give a contribution in the limit $L\to \infty$.

Let us look at the term linear in $\epsilon$ in (\ref{SPc}). One has
\begin{equation} 
\int_{-L}^L\epsilon(\xi) f(xs(\xi))d\xi=\frac{1}{\sqrt{2L}}\int_{0}^L\te\left(f(xs(\xi))-f(xs(-\xi))\right)d\xi
\end{equation}
By adding and subtracting $f(0)$ the previous integral can be written as
\begin{equation}
\frac{1}{\sqrt{2L}}\int_{0}^L\te\Big(f(xs(\xi))-f(0)\Big)d\xi-\int_{0}^L\te\Big(f(xs(-\xi))-f(0)\Big)d\xi
\end{equation}
If $f(x)$ is Lipschitz at $x=0$ then one has, for some constant $C$, $|f(xs(\xi))-f(0)|\leq C|xs(\xi)|$
\begin{equation} \begin{split}
&|\int_{-L}^L\epsilon(\xi) f(xs(\xi))d\xi|\leq \frac{C|x|}{\sqrt{2L}}\int_{0}^L \left(|s(\xi)-s(-\xi)|\right)d\xi
\end{split}\end{equation}
and since $s(\xi)$ is an $S^\infty$ function the integral is bounded for any $L$ and so the left hand side approaches  zero in the limit $L\to \infty$ $\square$.
\begin{cor}
The equation of motion (\ref{qpp}) reduces, in the limit $L\to\infty$, to
\begin{equation}\label{cont1}
\dot{Q}=P.
\end{equation}
\end{cor}
Indeed, it is sufficient to set $f(xs(\xi))=\exp(Q\eta(\xi)+\kappa(\xi))$ and use Proposition (\ref{procon}).

Let us now look at the equation of motion for $P$. One has
\begin{equation}
\dot{P}=1-\int_{-\infty}^{\infty}\eta(\xi)\exp(P\epsilon+Q\eta+\kappa)d\xi.
\end{equation}
We use the following Proposition
\begin{propn}\label{propc2}
Let $\eta(\xi)$ and $s(\xi)$ be two rapidly decreasing functions and consider the following expression:
\begin{equation}\label{SP3}
T(P,x,L)=\int_{-L}^L \eta(\xi)\exp(P\epsilon(\xi))f(xs(\xi))d\xi,
\end{equation}
where $f(x)$ is a continuous function of $x$ and $\epsilon(\xi)$ is given in (\ref{etaeps}). Then, for any finite values of $x$ one has
\begin{equation}\label{lim3}
\lim_{L\to\infty} T(P,x,L)=\int_{-\infty}^{\infty} \eta(\xi) f(xs(\xi))d\xi
\end{equation}
\end{propn}
\textbf{Proof}. For $\xi>\xi_0$, since $s(\xi)\to 0$ for $\xi \to \infty$, by continuity one has $|f(xs(\xi))- f(0)|<\delta(x)$ or $|f(xs(\xi))|<|f(0)|+\delta(x)$
So one has
\begin{equation}\begin{split}
&|\int_{-\infty}^{\infty} \eta(\xi) f(xs(\xi))d\xi|=|\int_{|t|<t_0} \eta(\xi) f(xs(\xi))d\xi+\int_{|t|\geq t_0} \eta(\xi) f(xs(\xi))d\xi|\leq \\
&\leq \int_{|t|<t_0} |\eta(\xi) f(xs(\xi))|d\xi+(|f(0)|+\delta(x))\int_{|t|\geq t_0} |\eta(\xi)|d\xi,
\end{split}\end{equation} 
 i.e. the integral $\int_{-\infty}^{\infty} \eta(\xi) f(xs(\xi))d\xi$ is finite since $\eta(\xi)$ is in $S^\infty$.
  It follows that if one considers the Taylor expansion of $\exp(P\epsilon)$ in the integral (\ref{SP3}), only the term proportional to $\epsilon^0$ survives in the limit $L\to \infty$ and we get (\ref{lim3}). $\square$

By using Proposition (\ref{propc2}) we get the following equations of motion for $P$ in the limit $L \to \infty$:
\begin{equation}\label{cont2}
\dot{P}=1-\int_{-\infty}^{\infty}\eta(\xi)\exp(Q\eta(\xi)+\kappa(\xi))d\xi.
\end{equation}
Since $\eta(\xi)$ is odd and $\kappa(\xi)$ is even, we can also write
\begin{equation}\label{cont3}
\dot{P}=1-\int_{-\infty}^{\infty}\eta(\xi)\exp(\kappa(\xi))\sinh(Q\eta(\xi))d\xi.
\end{equation}
The corresponding Hamiltonian is given by
\begin{equation}\label{evenHc}
\mathcal{H}=\frac{P^2}{2}-Q+\int_{-\infty}^{\infty}\exp(\kappa(\xi))\left(\cosh(Q\eta(\xi))-1\right)d\xi
\end{equation}

\subsection{The case of an odd Kernel}
Now we specify the functions $\epsilon$ and $\eta$ as:
\begin{equation}\label{etaepsodd}
\epsilon(x)=\frac{1}{\sqrt{2L}}\tilde{\epsilon}(x), \quad \eta(x)=\eta(-x)
\end{equation} 
where $\tilde{\epsilon}(x)^2=1$ for $x\in[-L,L]$, $\epsilon(x)=0$ for $x\notin [-L,L]$, $\te(x)$ even and $\eta_(x)$ even. The function $\kappa(x)$ must be odd. We assume further that
\begin{equation}\label{concon}
\lim_{L\to\infty}\int_{-L}^L\epsilon(\xi)d\xi=o(\sqrt{L}).
\end{equation}
For the equation of motion for $Q$ we use the following

\begin{propn}\label{prop4}
Let  $s(\xi)$ be a rapidly decreasing function and consider the following expression:
\begin{equation}\label{MPc}
M(P,x,L)=\int_{-L}^L \epsilon(\xi)\exp(P\epsilon(\xi))f(xs(\xi))d\xi,
\end{equation}
where $f(x)$ is a continuous function of $x$ and $\epsilon(\xi)$ is given in (\ref{etaepsodd}) and satisfies (\ref{concon}). Then, for any finite values of $x$ one has
\begin{equation}
\lim_{L\to\infty} M(P,x,L)=Pf(0).
\end{equation}
\end{propn}
\textbf{Proof}. By taking the Taylor series of the exponential in (\ref{MPc}) one has
\begin{equation}\label{tscon}
M(P,x,L)=\int_{-L}^L \epsilon(\xi)\left(1+P\epsilon(\xi)+\frac{P^2}{2}\epsilon(\xi)^2+\ldots\right)f(xs(\xi))d\xi.
\end{equation}
The term proportional to $\epsilon^2$ is:
\begin{equation}
P\int_{-L}^{L}\epsilon(\xi)^2f(xs(\xi))d\xi=\frac{1}{2L}\int_{-L}^L f(x s(\xi))d\xi.
\end{equation}
Exactly as for the equation (\ref{intf}), the previous expressions equals $Pf(0)$. It remains only the term linear in $\epsilon$ to look at, since all the powers of $\epsilon$ greater than $2$ in the Taylor expansion (\ref{tscon}) do not contribute in the limit $L\to \infty$. The linear term in $\epsilon$ is
\begin{equation}\label{eqline}
\int_{-L}^{L}\epsilon(\xi) f(xs(\xi))d\xi=\frac{1}{\sqrt{2L}}\int_{-L}^L \te(\xi) f(x s(\xi))d\xi.
\end{equation}
Due to the condition (\ref{concon}), since $f(x s(\xi)) \to f(0)$ as $\xi \to \infty$, the integral on the right hand side is $o(\frac{1}{\sqrt{L}})$ and it follows that the contribution from the term linear in $\epsilon$ in (\ref{MPc}) vanishes. Indeed the integral (\ref{eqline}) can be written as
\begin{equation}\label{eqline1}
\frac{1}{\sqrt{2L}}\int_{-L}^L \te(\xi) f(x s(\xi))d\xi=\frac{1}{\sqrt{2L}}\int_{-L}^L \te(\xi) \left(f(x s(\xi))-f(0)\right)d\xi+\frac{1}{\sqrt{2L}}\int_{-L}^L \te(\xi) f(0)d\xi.
\end{equation}
The last integral is equal to zero in the limit $L \to \infty$ due to condition (\ref{concon}). For the other integral, assume that $f(x)$ is Lipschitz at $x=0$. Then one has, for some constant $C$, $|f(xs(\xi))-f(0)|\leq C|xs(\xi)|$, giving
\begin{equation}\label{eqpro}
\frac{1}{\sqrt{2L}}|\int_{-L}^L \te(\xi) \left(f(x s(\xi))-f(0)\right)d\xi|\leq \frac{C|x|}{\sqrt{2L}}\int_{-L}^L |s(\xi)|d\xi
\end{equation}
and since $s(\xi)$ is a rapidly decreasing function the left hand side in (\ref{eqpro}) approaches to zero in the limit $L\to\infty$ $\square$.

\noindent
The equation of motion for $Q$ follows from the previous Proposition by taking $f(xs(\xi))=\exp(Q\eta(\xi)+\kappa(\xi))$. We get
\begin{equation}
\dot{Q}=P.
\end{equation}

For the equation of motion for $P$ we can use Proposition (\ref{propc2}), still valid if $\epsilon(\xi)$ is given as in (\ref{etaepsodd}). We get
\begin{equation}
\dot{P}=1-\int_{-\infty}^{\infty}\eta(\xi)\exp(Q\eta(\xi)+\kappa(\xi))d\xi.
\end{equation}
The function $\eta(\xi)$ is even and $\kappa(\xi)$ is odd, whereas the interval of summation is symmetric. It follows that only the even portion of $\exp(\kappa(\xi))$ survives, giving
\begin{equation}
\dot{P}=1-\int_{-\infty}^{\infty}\eta(\xi)\exp(Q\eta(\xi))\cosh(\kappa(\xi))d\xi.
\end{equation} 
The corresponding Hamiltonian is given by
\begin{equation}\label{oddHc}
\mathcal{H}=\frac{P^2}{2}-Q+\int_{-\infty}^{\infty}\exp(Q\eta(\xi))\left(\cosh(\kappa(\xi))-1\right)d\xi.
\end{equation}

The Remarks (\ref{rem1})-(\ref{rem2}) given for the discrete case can be easily translated to the continuous case.

\begin{rem}\label{rem3}
The equations of motion, both for an odd Kernel and on even one, are given by
\begin{equation}\label{eqmotremc}
\dot{Q}=P, \quad \dot{P}=1-\int_{-\infty}^{+\infty}\eta(\xi)\exp(Q\eta(\xi)+\kappa(\xi))d\xi,
\end{equation}
where the functions $\epsilon(x)$ and $\eta(x)$ have opposite parity with respect to $\kappa(x)$. The above system has always just one equilibrium position, say $(Q,P)=(Q_0,0)$, if the function $\eta(x)$ has not a definite sign. In this case the motion is bounded and the equilibrium position is a center.
\end{rem}
Again, the function defining $\dot{P}$ in (\ref{eqmotremc}) is a decreasing function of $Q$ as it follows immediately by taking its derivative with respect to $Q$. Further, for $Q\to -\infty$, $\dot{P} \to +\infty$ (we assume that there will be always a set of positive measure for which $\eta(x)$ is negative), whereas for $Q \to +\infty$, $\dot{P} \to -\infty$ (we assume that there will be always a set of positive measure for which $\eta(x)$ is positive), and one has just one zero. 
The motion must be bounded for the same reasons given in Remark (\ref{rem1}). Further, since (\ref{eqmotremc}) is Hamiltonian with the Hamiltonian given by
\begin{equation}
\mathcal{H}=T+V=\frac{P^2}{2}-Q+\int_{-\infty}^{+\infty}\left(\exp(Q\eta(\xi)+\kappa(\xi))-1\right)dx,
\end{equation} 
the potential $V$ has a global minimum in $Q_0$, as can be easily seen by taking its first and second derivative. Also it approaches $+\infty$ for $Q\to \pm \infty$ and is an infinite potential well.

\begin{rem}\label{rem3.1}
If the function $\eta(x)$ has just one sign, then the system (\ref{eqmotremc}) has one equilibrium position if $\eta(x)\geq 0$ and has no equilibrium position (open orbits) if $\eta(x) \leq 0$.
\end{rem}
The previous Remark follows by looking at the properties of the system (\ref{eqmotremc}) when $\eta(x)\geq 0$ or $\eta(x)\leq 0$, as did for the Remark (\ref{rem3}).

\begin{rem}\label{rem4}
The equations of motion (\ref{eqmotremc}) hold true also if the functions $\eta(x)$ and $\kappa(x)$ belong to $L^1$. The Propositions (\ref{procon}), (\ref{propc2}) and (\ref{prop4}) indeed hold also for $\eta(x)$ and $\kappa(x)$ belonging to $L^1$.
\end{rem}
An example of sequences in $L^1$ will be given in the subsection (\ref{subexc}). We end this Section giving a Remark, like in the discrete case, about the possibility to get a mixed odd/even Hamiltonian.
\begin{rem}
It is possible to take a flow that is a combination of the Hamiltonian containing the even Kernel $\mathcal{H}_e$ (\ref{evenHc}), and the Hamiltonian containing the odd Kernel $\mathcal{H}_o$ (\ref{oddHc}):
\begin{equation}
\alpha\mathcal{H}_e+(1-\alpha)\mathcal{H}_o=\frac{P^2}{2}-Q+\int_{-\infty}^{\infty}\left(\alpha e^{Q\eta^o(\xi)+\kappa^e(\xi)}+(1-\alpha)e^{Q\eta^e(\xi)+\kappa^o(\xi)}-1\right)d\xi.
\end{equation}
where the superscripts on the sequences remind if they are odd or even. If the functions $\eta^o(\xi)$ and $\eta^{e}(\xi)$  have not a definite sign, the potential
\begin{equation}
V=-Q+\int_{-\infty}^{\infty}\left(\alpha e^{Q\eta^o(\xi)+\kappa^e(\xi)}+(1-\alpha)e^{Q\eta^e(\xi)+\kappa^o(\xi)}-1\right)d\xi
\end{equation}
has just one minima for $\alpha\in [0,1]$ and is an infinite potential well. so that the corresponding motion is bounded and periodic. The proof being the same as in Remark (\ref{rem3}) and is omitted.
\end{rem}

In the next Section we will give some particular realization of the equations of motion, specifying the sequences $\eta_n$, $\epsilon_n$ and $\kappa_n$ for the discrete case or the functions $\eta(x)$, $\epsilon(x)$ and $\kappa(x)$ for the continuous case.

\section{Some examples}
In this Section we will give an analytic example of discrete model by starting from suitable $\epsilon_n$, $\eta_n$, $\kappa_n$. Successively, we will give an analytic example for the continuous model. Some numerics will be also given.

\subsection{An example for the discrete case}
We choose  the following sequences for $\epsilon_n$, $\eta_n$, $\kappa_n$\footnote{Given the choices of $\epsilon_n$ and $\eta_n$, we just made a particular choice of $\kappa_n$. This corresponds to fix the value of the Casimirs}:
\begin{equation}\label{cho}
\begin{split}
& \epsilon_n=\frac{1}{\sqrt{2N}}\te_n, \;\\
&\eta_n=\sqrt{\frac{e^{2a}-1}{2(1-e^{-2aN})}}e^{-a|n|}\te_n\\
&k_n=Ce^{-a|n|}
\end{split}
\end{equation}
where $\te_n$ is the sequence given in (\ref{cho1}). In the previous it is assumed that $\te_n$ has compact support in $[-N, N]$. Both $\epsilon_n$ and $\eta_n$ are normalized, in the $\ell^2$ norm, to the unity.

From the general formulae (\ref{dotqodd}) and (\ref{dotpodd}) we get
\begin{equation}
\dot{Q}=P, \quad \dot{P}=1-\sum_{n\in {\mathbb Z}}\eta_n\exp(Q\eta_n+\kappa_n).
\end{equation}
Let us look at the sum defining $\dot{P}$. By taking the Taylor series of the exponential we can write
\begin{equation}
\sum_{n\in {\mathbb Z}}\eta_n\exp(Q\eta_n+\kappa_n)=\sum_{n\in {\mathbb Z}}\sum_{k=0}^{\infty}\eta_n\frac{(Q\eta_n+\kappa_n)^k}{k!}.
\end{equation}
Now we can use the binomial expansion to give
\begin{equation}
\sum_{n\in {\mathbb Z}}\sum_{k=0}^{\infty}\eta_n\frac{(Q\eta_n+\kappa_n)^k}{k!}=\sum_{n\in {\mathbb Z}}\sum_{k=0}^{\infty}\sum_{m=0}^k{k \choose m}\frac{Q^m\eta^{m+1}_n\kappa^{k-m}_n}{k!}
\end{equation}
The sequence $\kappa_n$ is even, whereas $\eta_n$ is odd, so only when $m$ is odd one has a contribution different from zero, since the sum in $n$ is on a symmetric interval. So we can write
\begin{equation}
\sum_{n\in {\mathbb Z}}\sum_{k=0}^{\infty}\sum_{m=0}^k{k \choose m}\frac{Q^m\eta^{m+1}_n\kappa^{k-m}_n}{k!}=\sum_{n\in {\mathbb Z}}\sum_{k=0}^{\infty}\sum_{m=0}^k{k \choose m}\frac{(1-\cos(\pi m))}{2}\frac{Q^m\eta^{m+1}_n\kappa^{k-m}_n}{k!}.
\end{equation}
We now recall, from (\ref{cho}), that when $m+1$ is even $\eta^{m+1}_n=w^{m+1} e^{-a(m+1)|n|}$, where $w=\sqrt {\frac{e^{2a}-1}{2}}$. By considering also the explicit formula for $\kappa_n$ from (\ref{cho}) and by performing the sum over $n$ we can write
\begin{equation}
\sum_{n\in {\mathbb Z}}\sum_{k=0}^{\infty}\sum_{m=0}^k{k \choose m}\frac{Q^m\eta^{m+1}_n\kappa^{k-m}_n}{k!}=\sum_{k=0}^{\infty}\sum_{m=0}^k{k \choose m}(1-\cos(\pi m))\frac{Q^m w^{m+1}_n C^{k-m}}{k!(e^{a(k+1)}-1)}
\end{equation}
The sum over $m$ can be performed by using the binomial formula, we have
\begin{equation}\begin{split}
&\sum_{k=0}^{\infty}\sum_{m=0}^k{k \choose m}(1-\cos(\pi m))\frac{Q^m w^{m+1}_n C^{k-m}}{k!(e^{a(k+1)}-1)}=\\
&=w\sum_k\frac{(C+Qw)^k}{k!} \frac{1}{e^{a(k+1)}-1}-w\sum_k\frac{(C-Qw)^k}{k!}\frac{1}{e^{a(k+1)}-1}
\end{split}\end{equation}
By summarizing, the equation of motion for $\dot{P}$ is given by
\begin{equation}\label{thep}
\dot{P}=1-w\sum_k\frac{(C+Qw)^k}{k!} \frac{1}{e^{a(k+1)}-1}+w\sum_k\frac{(C-Qw)^k}{k!}\frac{1}{e^{a(k+1)}-1}
\end{equation}
We notice that the series (\ref{thep}) can be also expressed in terms of the following function:
\begin{equation}\label{defFz}
F(z)=\sum_{n=1}r^nz^{r^n}=z\frac{d}{dz}\sum_{n=1}(z^{r^n}-1),
\end{equation}
that is an entire function of $z$ in the complex plane for $r\in (0,1)$. Indeed, if we set $r=e^{-a}$, with $a>0$ we get the following equations of motion
\begin{equation}\label{eqm2}
\begin{split}
&\dot{Q}=P,\\
&\dot{P}=1+w(F(e^{C-Qw})-F(e^{C+Qw}))
\end{split}
\end{equation}
where the function $F(z)$ defined in (\ref{defFz}) satisfies the functional equation
\begin{equation}
rF(z^r)=F(z)-rz^r, 	\quad r\doteq e^{-a},
\end{equation}
Notice that the function $w(F(e^{C-Qw})-F(e^{C+Qw}))$ is \emph{decreasing} in $Q$ and decreases from $+\infty$ for $Q\to -\infty$ to $-\infty$ for $Q\to +\infty$, meaning that there is just one equilibrium configuration.

Alternatively but equivalently, the equations of motion (\ref{eqm2}) can be defined by the function $f(z)=F(\exp(z))$, given by the series
\begin{equation}\label{fz}
f(z)=\sum_{k=0} \frac{z^k}{k!}\frac{r^{k+1}}{1-r^{k+1}}.
\end{equation}
Indeed, by taking the finite series
\begin{equation}
\sum_{n=1}^N r^n\exp(zr^n),
\end{equation}
by expanding the exponential in its Taylor series, summing over $n$ and letting $N\to \infty$ one gets $F(\exp(z))=f(z)$, with $f(z)$ defined as in (\ref{fz}). 

The previous equations of motion of this infinite discrete system stem from the Hamiltonian function
\begin{equation}
H=\frac{P^2}{2}-Q+G(e^{C-Qw})-G(e^{C+Qw}),
\end{equation}
where the function $G(z)$ is defined by the series
\begin{equation}
G(z)=\sum_{n=1}(z^{r^n}-1)
\end{equation}
Equivalently, the Hamiltonian can be written in terms of the following function $g(z)$:
\begin{equation}
g(z)=\sum_{k=0} \frac{z^{k+1}}{(k+1)!}\frac{r^{k+1}}{1-r^{k+1}}.
\end{equation}
with
\begin{equation}
H=\frac{P^2}{2}-Q+g(C-Qw)-g(C+Qw).
\end{equation}
As for the function $f(z)$ and $F(z)$ above, the relation between $g(z)$ and $G(z)$ is simply given by $g(z)=G(\exp(z))$

\subsubsection{The numerics of the discrete case}
\begin{figure}[H]
\centering
\includegraphics[scale=0.65]{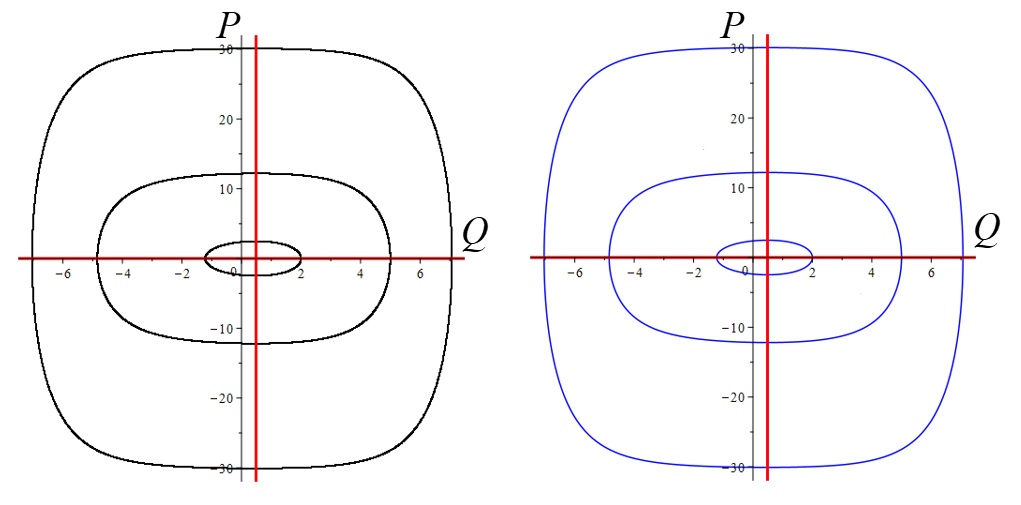}
\caption{Plot of three orbits of the system (\ref{numeq}): on the left there is the numeric, on the right the plot of the corresponding Hamiltonian level curves. In red the nullclines defined by $\dot{P}=0$ and $\dot{Q}=0$.}
\label{fig2}
\end{figure}

Let us set the parameter $w$ equal to 1 and $C$ also equal to 1. Since $w=\sqrt{\frac{e^{2a}-1}{2}}$, it follows that $a=\ln(3)/2$ and $r=\frac{1}{\sqrt{3}}$. The equations of motion can be written as
\begin{equation}\label{numeq}
\dot{Q}=P, \quad \dot{P}=1+f(1-Q)-f(1+Q)
\end{equation}
where 
\begin{equation}
f(z)=\sum_{k=0}^{\infty}\frac{z^k}{k!}\frac{1}{3^{\frac{k+1}{2}}-1}.
\end{equation}
By truncating the series at $n=N$ and noticing that $\frac{1}{3^{\frac{k+1}{2}}-1}\leq \frac{\sqrt{3}}{2}3^{-\frac{k}{2}}$ for $k>1$, one gets the following bound for the reminder:
\begin{equation}
|R_{N}|=|\sum_{k=N+1}^{\infty}\frac{z^k}{k!}\frac{1}{3^{\frac{k+1}{2}}-1}|\leq \frac{\sqrt{3}}{2}\sum_{k=N+1}^{\infty}\frac{(\frac{|z|}{\sqrt{3}})^k}{k!}=\frac{\sqrt{3}}{2}\frac{(\frac{|z|}{\sqrt{3}})^{N+1}}{(N+1)!}\exp(\frac{\theta|z|}{\sqrt{3}}),
\end{equation} 
where $\theta \in (0,1)$ and we used the Lagrange form of the reminder for the expansion of $\exp(\frac{|z|}{\sqrt{3}})$. Just to get an idea, for $N=30$ the previous formula gives an $R_{30}$ of the order  $10^{-8}$ for $z=10$ and $\theta=1$ and an $R_{40}$ of $10^{-16}$ for $N=40$ and the same values of $z$ and $\theta$, whereas the values of the sums from $k=0$ to $k=30$ and from $k=0$ to $k=40$ are about $2\cdot 10^2$. We use the series up to $N=40$ and integrate numerically the equations of motion with a standard Runge-Kutta-Fehlberg method with Maple. The Figure (\ref{fig2}) shows the numerical integration of the system (\ref{numeq}) with three different initial conditions: $(Q(0),P(0))=(10,0)$, $(Q(0),P(0))=(8,5)$ and $(Q(0),P(0))=(5,0)$. The corresponding values of the energies are $607.38102$, $207.09954$ and $33.39845$. The plots on the right, in blue, are the plots of the analytical curves corresponding to the level curves defined by the Hamiltonian in these three cases and coincide with the orbits. The curve in red is the curve defined by $\dot{P}=0$: the equilibrium position is where this curve intersect the $Q$ axis. 

\subsection{An example with $\eta_n$ and $\kappa_n$ in $\ell^1$ and $\kappa_n$ odd.}\label{subexd}
As specified in the Remark (\ref{rem2}), the equations of motion (\ref{eqmotrem}) holds true also if the sequences $\eta_n$ and $\kappa_n$ belong to $\ell^1$. The properties of the orbits given in the Remark (\ref{rem1}) or in the Remark (\ref{rem1.1}) are also valid in these cases. As an example of the differential equations that one obtains, we can take $\epsilon_n$ like in (\ref{cho2}) (hence $\kappa_n$ odd)
\begin{equation}\label{expld}
\begin{split}
&\eta_n=\frac{3\sigma}{\pi^2 n^2}, \; \; n\neq 0, \quad \eta_0=0,\quad \sigma^2=1\\
&\kappa_n=\frac{C\sgn(n)}{n^2}\; \; n\neq 0, \quad \kappa_0=0.
\end{split}
\end{equation}
From formula (\ref{eqmotrem}), to get $\dot{P}$ we need to evaluate the following sum
\begin{equation}
\sum_{n\in \mathbb{Z}}\eta_n\exp(Q\eta_n+\kappa_n),
\end{equation}
and, by using explicitly the relations (\ref{expld}) we get
\begin{equation}
\sum_{n\in \mathbb{Z}}\eta_n\exp(Q\eta_n+\kappa_n)=\frac{3\sigma}{\pi^2}\sum_{n=1}^{\infty}\frac{1}{n^2}\left(\exp\left(\frac{\frac{3\sigma}{\pi^2}Q+C}{n^2}\right)+\exp\left(\frac{\frac{3\sigma}{\pi^2}Q-C}{n^2}\right)\right).
\end{equation}
By taking the Taylor series of the exponential and defining the exponential generating function of the Riemann Zeta function on the even integers as
\begin{equation}
\zeta_{even}(z)=\sum_{k=0}^{\infty}\frac{z^k}{k!}\zeta(2k+2),
\end{equation}
we can write
\begin{equation}
\dot{Q}=P, \quad \dot{P}=1-\frac{3\sigma}{\pi^2}\zeta_{even}\left(\frac{\frac{3\sigma}{\pi^2}Q+C}{n^2}\right)-\frac{3\sigma}{\pi^2}\zeta_{even}\left(\frac{\frac{3\sigma}{\pi^2}Q-C}{n^2}\right)
\end{equation}
Notice that in this example the curve defining $\dot{P}=0$ has a different limit for $Q\to -\infty$ with respect to the general case discussed in the Remark (\ref{rem1}) and shown in Figure (\ref{fig1}): indeed, if $\sigma=1$, the $\eta_n$ are all non negative and hence in the limit $Q\to -\infty$ one has $\dot{P}=1$. The existence of just one equilibrium configuration is ensured by $\lim_{Q\to+\infty} \dot{P}=-\infty$ and from the monotonic decrease of $\dot{P}$. If $\sigma=-1$ then $\dot{P}$ is always positive and one has open orbits and no equilibrium configurations.

\subsection{An example for the continuous case}

We choose  for $\epsilon(x)$ the class of odd piecewise constant functions with a compact support in $(-L, L)$. The function $\eta(x)$ also is odd, whereas $\kappa(x)$ must be even. By requiring the $L^2$ normalization for $\epsilon$ and $\eta$ we can write ($a$ and $C$ are two parameters)
\begin{equation}\label{en}
\begin{split}
&\epsilon(x)=\frac{1}{\sqrt{2L}}\tilde{\epsilon}(x), \quad \tilde{\epsilon}^2=1. \\
&\eta(x)=\frac{a}{\sqrt{1-\exp(-2a^2 L)}}\exp(-a^2|x|)\tilde{\epsilon}(x),\\
&\kappa(x)=C\exp(-a^2|x|).
\end{split}
\end{equation}
Let us look at the equations of motion for $P$ (\ref{dotp}) and $Q$ (\ref{dotq}). From the general formulae (\ref{cont1}) and (\ref{cont2}) we get
\begin{equation}
\dot{Q}=P, \quad \dot{P}=1-\int_{-\infty}^{+\infty}\eta(\xi)\exp(Q\eta(\xi)+\kappa(\xi))d\xi
\end{equation}
We take the Taylor series expansion of the exponential in the equation for $\dot{P}$ and then use the binomial formula to get
\begin{equation}\label{initf1}
\int_{-\infty}^{+\infty}\eta(\xi)\exp(Q\eta(\xi)+\kappa(\xi))d\xi=\sum_{n=0}^{\infty}\sum_{j=0}^n {n \choose j}\int_{-\infty}^\infty d\xi \frac{Q^j\eta(\xi)^{j+1}\kappa(\xi)^{n-j}}{n!}
\end{equation}
Since $\eta$ is an odd function, whereas $\kappa$ is even, only when $j$ is odd there is a contribution different from zero. So, by taking into account the expressions (\ref{en}) and by noticing that $\tilde{\epsilon}^{j+1}=1$ for $j$ odd we can write:
\begin{equation}
\sum_{j=0}^n {n \choose j}\int_{-\infty}^\infty d\xi \frac{Q^j\eta(\xi)^{j+1}\kappa(\xi)^{n-j}}{n!}=\sum_{j=0}^n {n \choose j}\frac{1-\cos(\pi j)}{2}\int_{-\infty}^\infty d\xi \frac{Q^jC^{n-j}\alpha^{j+1}e^{-(n+1)\alpha^2|\xi|}}{n!}
\end{equation}
that is, by performing the integral
\begin{equation}\label{bfh}
\sum_{j=0}^n {n \choose j}\int_{-\infty}^\infty d\xi \frac{Q^j\eta(\xi)^{j+1}\kappa(\xi)^{n-j}}{n!}=\sum_{j=0}^n {n \choose j}\frac{\alpha^{j-1}Q^jC^{n-j}(1-\cos(\pi j))}{(n+1)!}.
\end{equation}
The sum over $j$ can be carried out by using the formula for the binomial expansion. The sum in equation (\ref{bfh}) is just given by
\begin{equation}
\frac{(C+Qa)^n}{a(n+1)!}-\frac{(C-Qa)^n}{a(n+1)!}.
\end{equation}
Going back to equation (\ref{initf1}) we can write
\begin{equation}
\int_{-\infty}^{+\infty}\eta(\xi)\exp(Q\eta(\xi)+\kappa(\xi))d\xi=\sum_{n=1}\left(\frac{(C+Qa)^n}{a(n+1)!}-\frac{(C-Qa)^n}{a(n+1)!}\right),
\end{equation}
that can be summed to
\begin{equation}
\frac{e^{C+a Q}-1}{a(C+aQ)}-\frac{e^{C-aQ}-1}{a(C-aQ)}
\end{equation}
giving the following equations of motion
\begin{equation}\label{eqconnum}
\dot{Q}=P, \quad \dot P=1-\frac{e^{C+aQ}-1}{a(C+aQ)}+\frac{e^{C-aQ}-1}{a(C-aQ)}.
\end{equation}
The equilibrium position are given by ($P=0, Q=Q_0)$, where $Q_0$ is the root of the transcendental equation:
\begin{equation}\label{eq1}
\frac{e^{C+aQ}-1}{a(C+aQ)}+\frac{e^{C-aQ}-1}{a(C-aQ)}=1.
\end{equation}
As expected, this equation has always just one real root. For the sake of simplicity, let us just set $a=1$. Then, both the functions $\frac{e^x-1}{x}$ and $\frac{e^{-x}-1}{x}$ are increasing, so that the left hand side of (\ref{eq1}), as a function of $Q$, is decreasing when $a=1$. Also, it decreases from $+\infty$ to $-\infty$, crossing the $Q$-line only once. By looking at the properties of the functions $\frac{e^x-1}{x}$ and $\frac{e^{-x}-1}{x}$ it is also possible to give more precise statements about the intervals on the $Q$-line containing the root. For example, if $C\in (0,C^{*})$ where $C^{*}=\frac{x^{*}}{2}$ and $x^{*}$ is the only real root of the equation $\frac{e^x-1}{x}=2$ ($x^{*}\sim 1.25643$), then $Q_0 \in (C,x^{*}-C)$. In general, if $C$ is large and negative, then $Q_0\sim C$, whereas if $C$ is positive then $Q_0 \in (0,x^{0})$, decreasing to $0$ when $C$ increases. Here $x^0$ corresponds to the root of (\ref{eq1}) when $C=0$, i.e. $x^0\sim 0.93082$.

This equilibrium position is stable and indeed is a center. In fact, the linearization of the dynamical system corresponding to the Hamiltonian (\ref{ham}) around $Q_0$ follows by letting $P=\delta P$ and $Q=Q_0+\delta Q$ with $\delta P$ and $\delta Q$ satisfying 
\begin{equation}\label{lin}
\dot{\delta Q}=\delta P, \quad \dot{\delta P}=G(Q_0)\delta Q,
\end{equation}
where $G(Q)$ is the derivative of the left hand side of (\ref{eq1}). We have shown however that the left hand side of (\ref{eq1}) is decreasing, so the solution of (\ref{lin}) is oscillatory.

\noindent
In this special case, in the limit $L \to \infty$ one gets for the quantities in (\ref{ABC1})-(\ref{ABC3}): ${\cal A}=0,~{\cal C}=1$, while ${\cal B}$ is given by the following more complicated expression:

$${\cal B}=1/(C-Q_0)^2[\exp(C-Q_0)(C-Q_0-1)+1]+1/(C+Q_0)^2[\exp(C+Q_0)(C+Q_0 -1)+1]$$

\noindent
 As expected, ${\cal B}$ coincides with $F^\prime(Q_0)$ (\ref{eq1}),  so that the linearized equations yield a  harmonic motion. 

\subsubsection{The numerics of the continuous case}
\begin{figure}[H]
\centering
\includegraphics[scale=0.65]{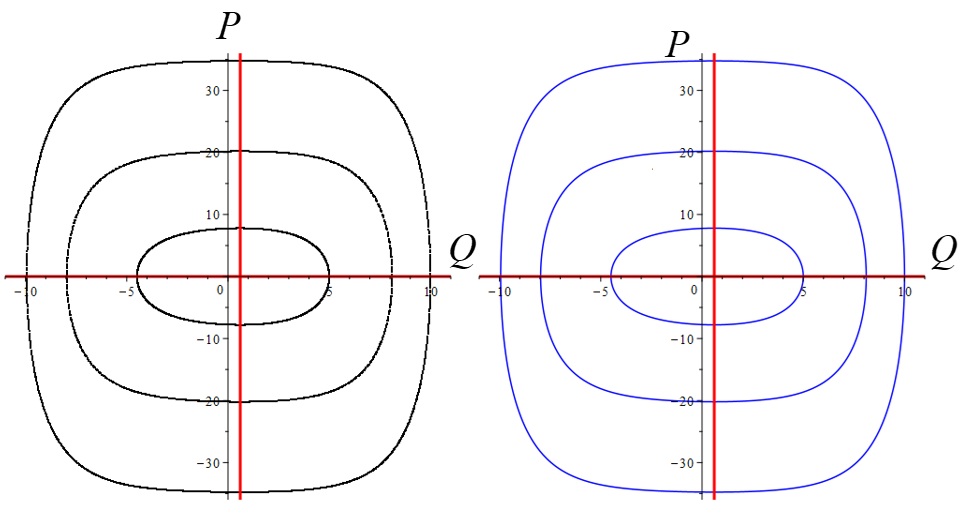}
\caption{Plot of three orbits of the system (\ref{eqconnum}): on the left there is the numeric, on the right the plot of the corresponding Hamiltonian level curves. In red the nullclines defined by $\dot{P}=0$ and $\dot{Q}=0$.}
\label{fig3}
\end{figure}

In order to numerically integrate the equation (\ref{eqconnum}), we set $\alpha=1$ and $C=1$. As for the discrete case, we plot three orbits. The orbits correspond to the following initial conditions: $(Q(0),P(0))=(2,0)$, $(Q(0),P(0))=(5,0)$ and $(Q(0),P(0))=(-7,5)$. The corresponding values of the energies are $6.61584$, $77.80793$ and $456.00834$. The plots on the right in Figure (\ref{fig3}), in blue, are the plots of the analytical curves corresponding to the level curves defined by the Hamiltonian in these three cases and coincide with the orbits. The curve in red is the curve defined by $\dot{P}=0$: the equilibrium position is where this curve intersect the $Q$ axis. 

\subsection{An example with $\eta(x)$ and $\kappa(x)$ in $L^1$ and $\kappa(x)$ odd.}\label{subexc}
As specified in Remark (\ref{rem4}), if the functions $\eta(x)$ and $\kappa(x)$ are in $L^1$, the equations of motion (\ref{eqmotremc}) still give a finite result. Let us take the following functions
\begin{equation}\label{explc}
\eta(x)=\frac{2\sigma}{\pi(1+x^2)}, \quad \sigma^2=1,\quad \kappa(x)=\frac{2}{\pi}\frac{C\sgn(x)}{(1+x^2)}.
\end{equation}
In order to deal with the equation for $\dot{P}$ in  (\ref{eqmotremc}), we use the following result: 
\begin{equation}\label{intBe}
\int_{0}^{\infty}\frac{1}{1+x^2}\exp(\frac{y}{1+x^2})dx=\frac{\pi}{2}\exp(y/2)I_{0}(y/2),
\end{equation}
where $I_0$ is the modified Bessel function of the first kind defined by
\begin{equation}
\frac{1}{\pi}\int_0^{\infty}\exp(z\cos(\xi))d\xi=I_0(z).
\end{equation}
Indeed, by the change of variables $x=\tan(\xi/2)$ (\ref{intBe}) becomes
\begin{equation}
\int_{0}^{\infty}\frac{1}{1+x^2}\exp(\frac{y}{1+x^2})dx=\frac{1}{2}\int_0^{\pi}\exp\left(\frac{y}{2}(1+\cos(\xi))\right)d\xi
\end{equation}
and formula (\ref{intBe}) follows. From the definitions (\ref{explc}) and formula (\ref{intBe}) we get for the equations of motion (\ref{eqmotremc})
\begin{equation}
\dot{Q}=P, \quad \dot{P}=1-\sigma\left(e^{\frac{\sigma Q+C}{\pi}}I_0\left(\frac{\sigma Q+C}{\pi}\right)+e^{\frac{\sigma Q-C}{\pi}}I_0\left(\frac{\sigma Q-C}{\pi}\right)\right)
\end{equation} 
We underline that in this example the curve defining $\dot{P}=0$ has a different limit for $Q\to -\infty$ with respect to the general case discussed in the Remark (\ref{rem3}). Indeed, if $\sigma=1$ then $\eta(x)$ is always positive and hence in the limit $Q\to -\infty$ one has $\dot{P}=1$. The existence of just one equilibrium configuration is ensured by $\lim_{Q\to+\infty} \dot{P}=-\infty$ and from the monotonic decrease of $\dot{P}$. If $\sigma=-1$ then $\dot{P}$ is always positive and one has open orbits and no equilibrium configurations.

\section{Concluding remarks}

In this paper we proposed an extension of the integrable Volterra model to the case of infinitely many species, either countable or uncountable. The corresponding models have been shown to possess the property of being maximally superintegrable since reduce to a system with just one degree of freedom. This property is independent on the choice of the parameters, but is essentially related to the fact that, in the integrable case, the interaction matrix $A$ has always rank 2. The one dimensional equations of motion are (\ref{Pdiscr})-(\ref{Qdiscr}) for the discrete case and (\ref{dotp})-(\ref{dotq}) for the continuous case. Clearly, it is necessary that the sequences $\epsilon_n$, $\eta_n$ and $\kappa_n$ for the discrete case or the functions $\epsilon(x)$, $\eta(x)$ and $\kappa(x)$ for the continuous case are such that the corresponding equations converge.

In our opinion, though the generalizations proposed are non trivial, they do not lead to an enrichment of  the dynamics of the system. Indeed, within the class  of sequences or continuous functions we have considered, in the infinite limit, both discrete and continuous, the behaviour of the system  looks even simpler than the one characterizing the case of a finite number of species. We may assert that  the maximal superintegrability, already discovered in \cite{RZ}, gets  in a sense strengthened by the infinite limit, inasmuch as there exists a general formula showing that, in canonical coordinates, the Hamiltonian takes, in a way unexpectedly, the classical one body feature ${\mathcal H} = \frac{P^2}{2} + V(Q)$. For the discrete case this is true if the sequence $\epsilon_n$ is chosen like in (\ref{cho1}) or (\ref{cho2}), i.e. is a set of $+1$ and $-1$, normalized in $\ell^2$, distributed evenly or oddly on the real line, whereas for the continuous case we take a normalized step-like piecewise constant function, evenly or oddly distributed on the real line. In order to get the convergence, the sequences $\eta_n$ and $\kappa_n$ can belong to the set of rapidly decreasing sequences $S^{\infty}$ or to the larger space $\ell^1$, and the corresponding equivalent assumptions must be made about the functions $\eta (x)$ and $\kappa (x)$ in the continuous case.

The explicit form the potential $V(Q)$ takes depends on the particular choices of the sequences $\eta_n$, $\kappa_n$ (or $\eta(x)$ and $\kappa(x)$) for the continuous case. We give some examples. Also, we have shown that with our particular choices of the growth rates $\epsilon_n$ ($\epsilon(x)$), the Hamiltonian system ${\mathcal H} = \frac{P^2}{2} + V(Q)$ possesses bounded periodic orbits if the sequence $\eta_n$ has not a fixed sign or if $\eta_n \geq 0$ for all $n$, whereas possesses open orbits if $\eta_n \leq 0$ for all $n$. Examples of  all these cases have been given. The same statements are true in the continuous case by requiring respectively that the function $\eta(x)$ has not a fixed sign, is non-positive or is non-negative.

We have also shown that taking an even or odd Kernel (or growth rates) is not limitative since it is possible to take a combination of the even and odd Hamiltonian flow and get a mixed Hamiltonian. The properties of the corresponding orbits are also been given.

The above results are strongly related to the family of sequences for the growth coefficients $\epsilon_n$ or the family  of growth function $\epsilon(x)$ chosen. In principle, a different choice of these coefficients or function can destroy the simple Hamiltonian structure described in this work. The instances here given are, in our opinion, significant for the following reasons: they give, in the infinite limit, both in the discrete and in the  continuous case, a finite and specific Hamiltonian model. Also, the uncountable case can be seen as the natural limit of the denumerable one. Further, the sequences $\eta_n$ and $\kappa_n$, or the functions $\eta(x)$ and $\kappa(x)$ for the continuous case, are essentially arbitrary (apart from the fact of  belonging to a suitable summable or integrable space) and the properties of the corresponding Hamiltonian models are universal, in the sense that, fixed the sequences $\eta_n$ and $\kappa_n$ or the functions $\eta(x)$ and $\kappa(x)$, the behaviour of the orbits are fixed (either periodic or open for specific cases). 

Finally, the connection between the growth coefficients and the automatic sequences, like the Rudin-Shapiro sequence, briefly mentioned in the text, is somehow unexpected and surely deserves much attention in future works. 

As for the unanswered questions: we have not been able to find an explicit formula for the period of the orbits, apart for the linearized case. For a finite number of species we have shown \cite{SRTZ} that, in suitable cases, the period of the linearized system is a very good approximation of the period of the orbits of the non linearized model. We expect that, if the nonlinearity is not too strong, the observation remains true also in our examples, but we didn't give any quantitative statement about this. Finally, it would be interesting to expand the class of sequences or functions for $\epsilon_n$ and $\epsilon(x)$ here given to other families: as said, the Hamiltonian structure that we obtained may be destroyed and richer models, for example with a greater number of equilibrium positions and different type of behaviours for the orbits, could be found.

\vspace{2mm}
\begin{center} {\bf Acknowledgments} \end{center}
F.Z. wishes to acknowledge the support of Universit\`a degli Studi di Brescia; INFN, Gr. IV - Mathematical Methods in NonLinear Physics and ISNMP - International Society of Nonlinear Mathematical Physics. O.R. and F.Z. wish to acknowledge the support of GNFM-INdAM.

\label{lastpage}
\vfill

\begin{thebibliography}{10}
\small
\bibitem{AS} J. P. Allouche, J. Shallit, {\it Automatic Sequences. Theory, Applications, Generalizations}, Cambridge University Press, Cambridge, (2003). 
\bibitem{Baigent9} S. Baigent, {\it Lotka-Volterra Dynamical Systems}, Dynamical and Complex Systems, LTCC Advanced Mathematics Series, 5, pp. 157-188, World Scientific, (2017).
\bibitem{Bountis11} T. Bountis, Z. Zhunussova, K. Dosmagulova, G. Kanellopoulos, {\it Integrable and non-integrable Lotka-Volterra systems}, Physics Letters A, Volume 402, (2021).
\bibitem{CDE15} S. A. Charalambides, P. A. Damianou, C. A. Evripidou, {\it On generalized Volterra systems}, Journal of Geometry and Physics, Volume 87, pp. 86-105, (2015).
\bibitem{Hone17} H. Christodoulidi , A.N.W. Hone, T. E. Kouloukas, {\it A new class of integrable Lotka-Volterra systems}, Journal of Computational Dynamics, Volume 6, Issue 2, pp. 223-237, (2019).
\bibitem{fernandez6} R. L. Fernandes and W. M. Oliva, {\it Hamiltonian dynamics of the Lotka-Volterra equations}, in International Conference on Differential Equations (Lisboa, 1995). pp. 327-334, World Sci. Publ., River Edge, NJ. (1998).
\bibitem{Goodwin12} R. M. Goodwin, {\it A growth Cycle}, in C.H. Feinstein, Editor, Socialism, Capitalism and Economic Growth, Cambridge University Press, (1967).
\bibitem{Hofba14} J. S. K. Hofbauer, {\it Dynamical Systems and Lotka-Volterra Equations}. Evolutionary Games and Population Dynamics. New York: Cambridge University Press, pp. 1-54, ISBN 0-521-62570-X, (1998).
\bibitem{KVM} M. Kac, P. Van Moerbeke, {\it{On an explicitly soluble system of nonlinear differential equations related to certain Toda lattices}}, Advances in Mathematics, 16 (2), pp. 160-169, (1975).
\bibitem{KF} A. N. Kolmogorov, S. V. Fomin, {\it{Elements of the Theory of Functions and Functional Analysis}}, Dover, New York, (1999).
\bibitem{Leigh10} E. R. Leigh, {\it The ecological role of Volterra's equations} Some Mathematical Problems in Biology. A modern discussion using Hudson's Bay Company data on lynx and hares in Canada from 1847 to 1903, (1968). 
\bibitem{L1} A. J. Lotka, {\it{Contribution to the Theory of Periodic Reactions}}, J.Phys.Chem. 14(3), pp. 271-274, (1910).
\bibitem{L2} A. J. Lotka, {\it{Analytic Note on Certain Rhythmic Relations in Organic Systems}}, Proc.Nat.Acad. 6, pp. 410-415 (1920).
\bibitem{MR} J. E. Marsden, T. S. Ratiu,{\it {Reduction of Poisson manifolds}}, Lett. Math. Phys. 11, pp. 161-169, (1986).
\bibitem{Peixe8} T. J. L. Peixe, {\it Lotka-Volterra Systems and Polymatrix Replicators}, Ph.D. Thesis, Universidade de Lisboa, (2015). 
\bibitem{RZ} O. Ragnisco, F. Zullo, {\it {The N-species integrable Volterra system as a maximally superintegrable Hamiltonian system}}, Open Communications in Nonlinear Mathematical Physics, ]ocnmp[ Vol.5, pp. 36-56, (2025).
\bibitem{SRTZ} M. Scalia, O. Ragnisco, B. Tirozzi, F. Zullo, {\it The Volterra Integrable case. Novel analytical and numerical results}, Open Communications in Nonlinear Mathematical Physics ]ocnmp[ Vol.4, pp. 188-211, (2024).
\bibitem{Scalia16} M. Scalia, A. Angelini, F. Farioli, G. F. Mattioli, O.Ragnisco, M. Saviano, {\it An Ecology and Economy coupling model}. A global stationary state model for a sustainable economy in the Hamiltonian formalism, Ecological Economics, Elsevier, 172, 106497, (2020).
\bibitem{Teschl} G. Teschl, {\it Almost Everything You Always Wanted to Know About the Toda Equation} Jahresber. Deutsch. Math.-Verein. 103, no. 4, pp. 149-162, (2001).
\bibitem{Toda} M. Toda, {\it Theory of Nonlinear Lattices}, Springer Series in Solids State Sciences 20, Berlin, (1978).
\bibitem{VanDerKamp13} P.H. Van der Kamp, D.I. McLaren, G.R.W. Quispel, {\it On a quadratic Poisson algebra and integrable Lotka-Volterra systems, with solutions in terms of Lambert's W function}, Regul. Chaot. Dyn., 30, pp. 382-407, (2025).
\bibitem{Griffin7} M. Visomirski, C. Griffin, {\it Integrability of generalised skew-symmetric replicator equations via graph embeddings}, J. Phys. A: Math. Theor. 58, 015701, (2025).
\bibitem{V1} V. Volterra, {\it{Lecons sur la Th\' eorie Math\' ematiques de la Lutte pour la Vie}}, Gauthiers Villars, Paris (1931).
\bibitem{Volterra1937} V. Volterra, {\it Principes de Biologie Math\' ematiques}, Acta biotheoretica Leiden, 3, pp. 6-39, (1937).

\end{thebibliography}
\end{document}